\documentclass[twocolumn,english]{article}
\usepackage[T1]{fontenc}
\usepackage{amssymb}
\usepackage{amsmath}
\usepackage{graphicx}
\usepackage{babel}
\usepackage{lscape}
\makeatletter
\usepackage{a4}
\input{epsfig.sty}
\def\fnum@table{\tablename~{\bf\thetable}}
\def\fnum@figure{\figurename~{\bf\thefigure}}
\def\tablename{\footnotesize{\bf Table}}
\def\figurename{\footnotesize{\bf Figure}}

\def\be{\begin{equation}}
\def\ee{\end{equation}}

\makeatother

\usepackage{babel}
\begin{document}

\title{\textbf{QGSJET-III model of high energy hadronic interactions:   The formalism}}

\author{Sergey Ostapchenko\\
\textit{\small Universit\"at Hamburg, II Institut f\"ur Theoretische
Physik, 22761 Hamburg, Germany}\\
}

\maketitle
\begin{center}
\textbf{Abstract}
\par\end{center}
The physics content of the QGSJET-III Monte Carlo generator 
of high energy hadronic collisions is described. In particular, a 
phenomenological implementation of higher twist corrections to hard parton
scattering processes is discussed in some detail. Additionally
 addressed is the
treatment of the so-called ``color fluctuation'' effects related to a
decomposition of hadron wave functions into a number of Fock states
characterized by different spatial sizes and different parton densities.
Selected model results regarding the energy-dependence of the total, elastic,
and diffractive proton-proton  cross sections are presented.

\section{Introduction\label{intro.sec}}

Nowadays high energy experiments both in the collider and cosmic ray (CR) fields
imply an extensive use of Monte Carlo (MC) generators of hadronic interactions.
At colliders, the primary goal of such generators is to describe Standard
Model backgrounds for new physics searches and to confront novel theoretical
ideas to experimental data. On the other hand, in the CR field, MC models of
high energy interactions play an important role, when interpreting experimental 
data. This is particularly so for investigations of very high energy CRs,
which are performed by indirect methods: studying various characteristics of
  extensive air showers (EAS) -- huge nuclear-electromagnetic
cascades induced by interactions of primary CR particles in the atmosphere
of the Earth -- and reconstructing the properties of those particles, based on
the measured EAS characteristics. Such applications imply a number of
requirements to CR interaction models. Those should be able to provide a
reasonable description of collisions with nuclei of various hadron species,
primarily, of (anti)nucleons, pions, and kaons, over a wide energy range:
from fixed target energies up to some $10^{12}$ GeV laboratory  energy.
Additionally, the steeply falling down primary CR flux and the cascade nature
of extensive air showers enhance the importance
of forward secondary particle production. Last but nor least, given the 
scarcity of available experimental data regarding such forward production 
and the lack of possibility to re-tune  MC generators, based on CR data, 
a substantial predictive power is required from CR interaction models.

Over the past three decades, the QGSJET \cite{kal93,kal94,kal97} and
QGSJET-II \cite{ost06,ost11,ost13} MC generators proved to be very successful 
regarding the analyses and interpretations of various  CR data,
notably, from air shower experiments. In the current work, we report a further
development of the model framework, related to taking into consideration
the so-called dynamical higher twist corrections to hard parton scattering
processes and to the implementation of  ``color fluctuation''
 effects in high energy
hadronic collisions. While the treatment of secondary particle production and
the application of the model to calculations of EAS characteristics will be
discussed elsewhere \cite{ost23}, we concentrate here on the description of the
model formalism, providing also some selected results regarding the 
energy-dependence of the total, elastic, and diffractive proton-proton
 cross sections.

The paper is organized as follows. In Section \ref{rft.sec}, we discuss the 
 Reggeon Field Theory (RFT) approach to multiple scattering in hadronic 
 collisions. Section \ref{hard-pom.sec} is devoted to the treatment of
 hard parton scattering within the  RFT framework. In Section  \ref{gw.sec},
 we address the implementation of  color fluctuation effects.
 The treatment of nonlinear interaction effects due to Pomeron-Pomeron 
 interactions is described in Section \ref{enhanced.sec}. Section \ref{ht.sec} is devoted to phenomenological implementation of dynamical power corrections to hard parton
 scattering. The model generalization to the case of nuclear collisions
 and the MC realization of the formalism are described in Section \ref{MC.sec}.
 In  Section \ref{results.sec}, we present and discuss selected model results.
 Finally, we conclude in  Section \ref{concl.sec}.

\section{Multiple scattering in the Reggeon Field Theory  \label{rft.sec}}
High energy hadronic collisions are predominantly multiple scattering 
processes, 
being mediated by multiple parton cascades developing between the interacting projectile
and target hadrons (nuclei). While a perturbative description of multiple scattering
(so-called multi-parton interactions) is an actively developing field (see, e.g., 
\cite{die12} for a review), the corresponding treatment in  MC generators
has to rely presently on the old  RFT formalism \cite{gri68}.

Since the underlying, so to say  ``elementary'', parton cascades develop,
 at least partly,
in the nonperturbative domain of low parton virtualities, 
where the notion of partons
can be used for a qualitative discussion only,
 one is forced to rely on an effective
macroscopic description for such cascades -- treating them as Pomeron exchanges.
The Pomeron exchange eikonal (the imaginary part\footnote{The real part of the Pomeron
amplitude can be neglected in the high energy limit.}
 of the corresponding amplitude) is usually chosen in the form
\begin{eqnarray}
\chi_{hp}^{\mathbb P}(s,b)=\frac{\gamma_h\,\gamma_p\,
(s/s_0)^{\alpha_{\mathbb P}(0)-1}}{R_h^2+R_p^2+
\alpha_{\mathbb P}'(0)\,\ln (s/s_0)}\,
&& \nonumber \\
\times \; \exp \!\left[- \frac{b^2/4}{R_h^2+R_p^2+\alpha_{\mathbb P}'(0)\,\ln (s/s_0)}\right], &&
\label{eq:chi-pom}
\end{eqnarray}
where $\alpha_{\mathbb P}(0)$ and $\alpha_{\mathbb P}'(0)$ are, respectively, 
the intercept and the slope of the Pomeron Regge trajectory,
 $\gamma_h$ is the residue and $R_h^2$ the
slope for the Pomeron coupling to hadron $h$, 
$s$ and $b$ are,  correspondingly, 
the center-of-mass (c.m.) energy squared and impact parameter
for the collision,
and $s_0 \simeq 1$ GeV$^2$ -- the hadronic
mass scale.

\begin{figure}[htb]
\begin{centering}
\includegraphics[width=0.45\textwidth,height=3cm]{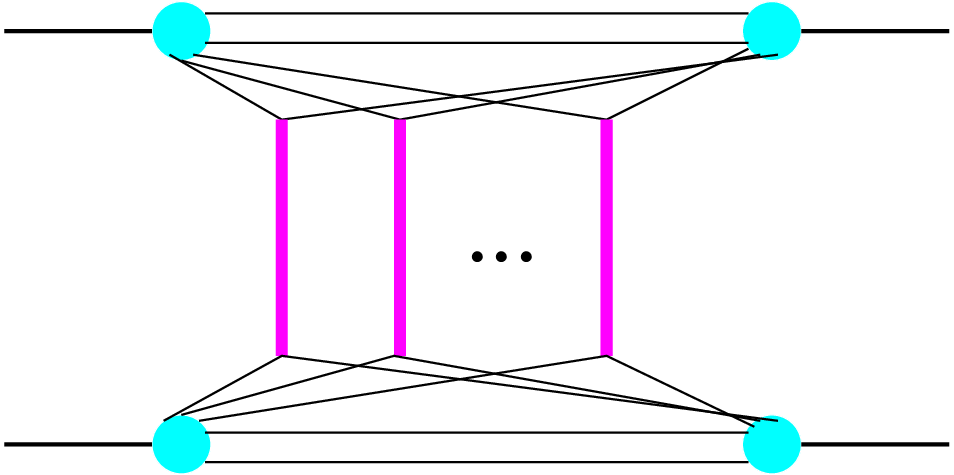}
\par\end{centering}
\caption{General multi-Pomeron contribution to hadron-hadron scattering amplitude;
elementary scattering processes (vertical thick lines) are described
as Pomeron exchanges.\label{Fig:mult}}
\end{figure}

Using eikonal description for multiple Pomeron emission vertices,
 one obtains the well-known simple
expressions for the total and elastic hadron-proton cross sections:
\begin{eqnarray}
\sigma^{\rm tot}_{hp}(s) = 2\int \! d^2b \left[1-e^{-\chi_{hp}^{\mathbb P}(s,b)}\right]
\label{eq:sig-eik-tot} &&\\
\sigma^{\rm el}_{pp}(s) = \int \! d^2b \left[1-e^{-\chi_{hp}^{\mathbb P}(s,b)}\right]^2.  &&
\label{eq:sig-eik-el}
\end{eqnarray}
Moreover, considering unitarity cuts of the corresponding elastic scattering diagrams
shown schematically in Fig.\ \ref{Fig:mult}
 and applying the so-called Abramovskii-Gribov-Kancheli (AGK)
cutting rules \cite{agk74}, one is able to obtain partial cross sections for various
inelastic final states corresponding to having precisely $n$  ``elementary'' production
processes ($n$ cut Pomerons):
\begin{equation}
\sigma^{(n)}_{hp}(s) = \int \! d^2b \; \frac{\left[2\chi_{hp}^{\mathbb P}(s,b)\right]^n}{n!}\; e^{-2\chi_{hp}^{\mathbb P}(s,b)}.\label{eq:sig(n)}
\end{equation}

To describe secondary particle production, one assumes that each cut Pomeron corresponds
to a creation of a pair of strings of color field, stretched between constituent partons [(anti)quarks or (anti)diquarks] 
of the interacting hadrons and models the breakup of those
strings by means of suitable string fragmentation procedures \cite{kai82,cap91}.
Importantly, the corresponding parameters can be expressed via intercepts of secondary
Regge trajectories \cite{kai87}.

\section{Treating hard scattering within the  RFT framework \label{hard-pom.sec}}
\begin{figure}[htb]
\begin{centering}
\includegraphics[width=0.45\textwidth,height=3cm]{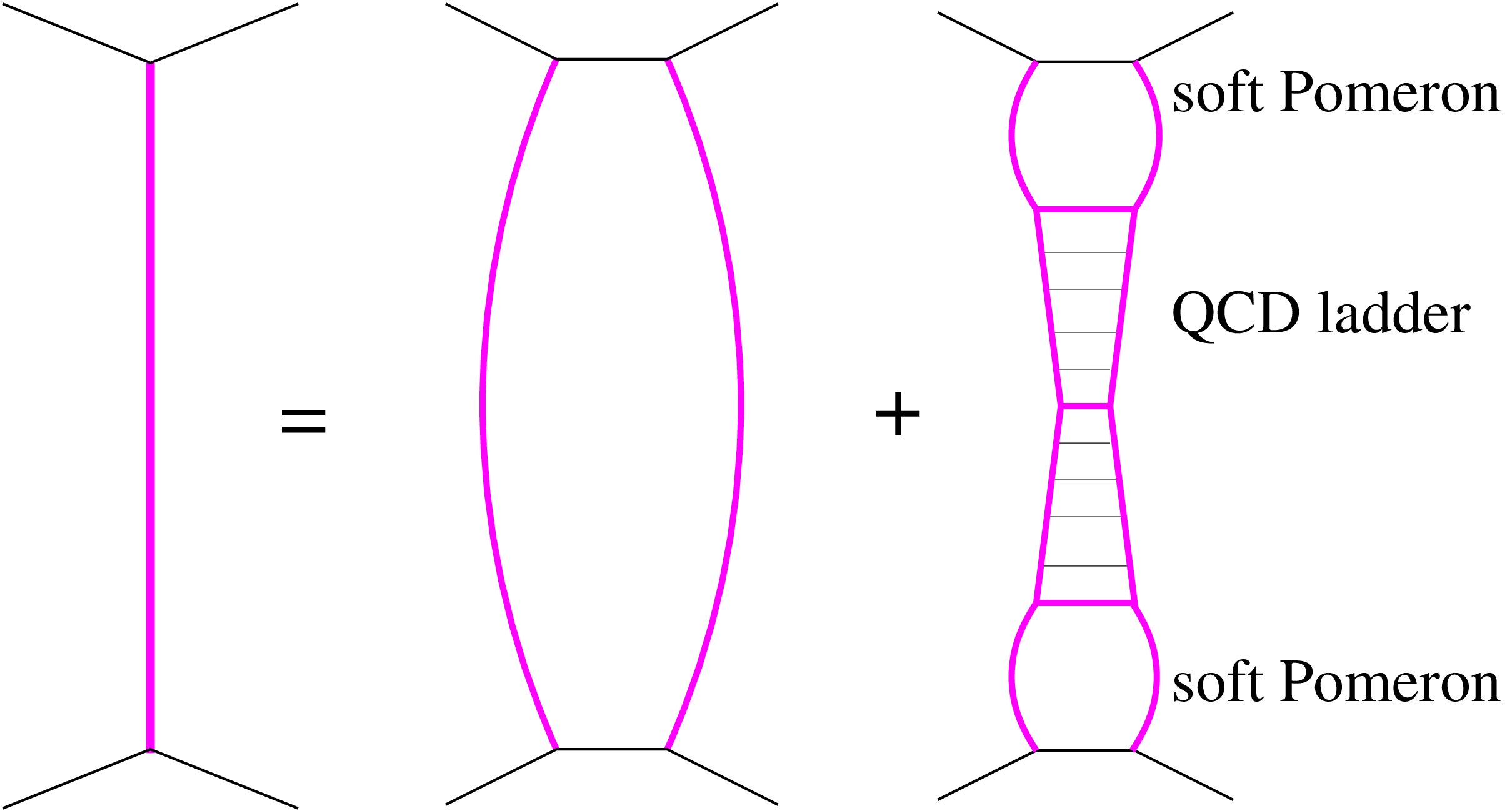}
\par\end{centering}
\caption{A ``general Pomeron'' (left-hand side) consists of the soft and semihard ones -- 
correspondingly the first and the second contributions in the  right-hand side.
\label{Fig:sh-pom}}
\end{figure}

While the original Gribov's formulation of RFT
 relied on the assumption that the bulk of hadron production is
characterized by small transverse momenta \cite{gri68},
 $p_t\lesssim 1$ GeV, the contribution of the so-called semihard processes corresponding to parton cascades developing, at least partly,
in the high $p_t$ domain becomes increasingly important in the very high energy limit. Indeed, the smallness of the corresponding strong coupling, $\alpha_s(p_t^2)$, in such cascades becomes compensated by large collinear and infrared logarithms and by a 
high parton density \cite{glr}. To treat such processes within the RFT framework,
the so-called ``semihard Pomeron'' approach has been proposed \cite{kal94,dre99,dre01,ost02}
(see also \cite{wer23} for a recent discussion). 
The underlying basic idea was to employ
the above-discussed phenomenological Pomeron description for purely
``soft'' nonperturbative (parts of) parton cascades: for parton virtualities
$|q^2|<Q_0^2$, while treating parton evolution in the perturbative $|q^2|>Q_0^2$
domain by means of the Dokschitzer-Gribov-Lipatov-Altarelli-Parisi (DGLAP) formalism \cite{gri72,alt77,dok77},
 with $Q_0^2$ being some chosen virtuality cutoff for the perturbative
quantum chromodynamics (pQCD) to be applicable. This allowed one to develop a Pomeron
calculus, based on a ``general Pomeron'', the latter being a sum of the soft and semihard ones, as shown symbolically in   Fig.\ \ref{Fig:sh-pom}.

In particular, Eqs.\ (\ref{eq:sig-eik-tot}--\ref{eq:sig(n)}) 
hold upon defining the eikonal $\chi^{\mathbb P}_{hp}$
as a sum $\chi^{{\mathbb P}_{\rm soft}}_{hp}+\chi^{{\mathbb P}_{\rm sh}}_{hp}$,
where $\chi^{{\mathbb P}_{\rm soft}}_{hp}$ is given by Eq.\ (\ref{eq:chi-pom}),
 while for
$\chi^{{\mathbb P}_{\rm sh}}_{hp}$ one obtains \cite{dre01,ost02}:
\begin{eqnarray}
\chi_{hp}^{\mathbb{P}_{{\rm sh}}}(s,b)=\frac{1}{2}\sum_{I,J}
\int\!\! d^{2}b'\!\!\int\!\frac{dx^{+}}{x^{+}}\frac{dx^{-}}{x^{-}}  &&\nonumber \\
\times\;
\chi_{Ih}^{\mathbb{P}_{{\rm soft}}}(\frac{s_{0}}{x^{+}},b')
\,\chi_{Jp}^{\mathbb{P}_{{\rm soft}}}(\frac{s_{0}}{x^{-}},|\vec{b}-\vec{b}'|)
 &&\nonumber \\
\times\;\sigma_{IJ}^{{\rm QCD}}(x^{+}\, x^{-}\,s,Q_{0}^{2},Q_{0}^{2})\,.
\label{chi-sh}&&
\end{eqnarray} 
Here $\sigma_{IJ}^{{\rm QCD}}(\hat{s},q_1^{2},q_2^{2})$ corresponds to the 
contribution of the DGLAP parton ``ladder'', with the
ladder leg partons $I$ and $J$ [(anti)quarks or gluons\footnote{We shall not discuss 
explicitly the contribution of hard interactions of valence quarks;
 see, e.g., \cite{dre01}
for the corresponding details.}] characterized by the virtualities  $q_1^2$
and   $q_2^2$, respectively:
\begin{eqnarray}
\sigma_{IJ}^{{\rm QCD}}(\hat{s},q_1^{2},q_2^{2})=
K\sum_{I',J'}\!\int\! dz^{+}dz^{-}\int\! dp_{t}^{2}&&\nonumber \\
\times \; E_{II'}^{{\rm QCD}}(z^{+},q_1^{2},\mu_{{\rm F}}^{2})\, 
E_{JJ'}^{{\rm QCD}}(z^{-},q_2^{2},\mu_{{\rm F}}^{2})&&\nonumber \\
\times\;\frac{d\sigma_{I'J'}^{2\rightarrow2}(z^{+}z^{-}\hat{s},p_{t}^{2},\mu_{\rm R})}
{dp_{t}^{2}} &&\nonumber \\
\times\;\Theta(\mu_{F}^2-\max [q_1^{2},q_2^{2}])\,,\label{sigma-hard}&&
\end{eqnarray}
where $d\sigma_{IJ}^{2\rightarrow2}/dp_{t}^{2}$ is the Born
parton cross section, $p_{t}$ being the parton transverse
momentum in the hard process, $\mu_{F}$ and  $\mu_{R}$ are the factorization 
and renormalization scales, respectively
(we use $\mu_{F}=\mu_{R}=p_{t}/2$), and the factor $K=1.5$ takes effectively
into account higher order QCD corrections. 
$E_{II'}^{{\rm QCD}}(z,q^{2},\mu_{{\rm F}}^{2})$
describes parton density evolution from the scale 
$q^{2}$ to $\mu_{{\rm F}}^{2}$,
subject to the initial condition 
$E_{II'}^{{\rm QCD}}(z,q^{2},q^{2})=\delta_I^{I'}\,\delta (1-z)$.

 In turn, the eikonal $\chi^{{\mathbb P}_{\rm soft}}_{Ih}$ corresponding
  to a soft Pomeron exchange between hadron $h$ and parton $I$
 is obtained from Eq.\ (\ref{eq:chi-pom}),  neglecting
the small slope of the Pomeron-parton coupling $R_{I}^{2}\sim 1/Q_{0}^{2}$
and replacing the vertex $\gamma_{p}$ by a parameterized
Pomeron-parton vertex  $V^{\mathbb P}_{I/h}$:
\begin{eqnarray}
\chi_{Ih}^{\mathbb{P}_{{\rm soft}}}(\hat{s},b) =\frac{\gamma_{h}\,
V^{\mathbb P}_{I/h}(s_{0}/\hat{s})\,
(\hat{s}/s_{0})^{\alpha_{\mathbb P}(0)-1}}
{R_{h}^{2}+\alpha_{\mathbb{P}}'(0)\,\ln(\hat{s}/s_{0})}&&\nonumber \\
\times \;\exp\!\left[-\frac{b^{2}/4}{R_{h}^{2}
+\alpha_{\mathbb{P}}'(0)\,\ln(\hat{s}/s_{0})}\right].\label{chi_aI}&&
\end{eqnarray}
We use\footnote{The ansatz, Eqs.\ (\ref{gamma-g}-\ref{gamma-q}), differs from the
one used in  \cite{dre01,ost02} by the factor $(1+x)^{b_h}$ chosen to
improve the large $x$ behavior of gluon PDFs.
More sophisticated parameterizations may generally be used for $V^{\mathbb P}_{I/h}$.} \cite{ost02}
\begin{eqnarray}
V^{\mathbb P}_{g/h}(x)=r_{g/\mathbb{P}}\,(1-w_{qg})\,
(1-x)^{\beta_{g/h}} &&\nonumber \\
\times \;(1+x)^{b_h}
\label{gamma-g}&&\\
V^{\mathbb P}_{q/h}(x)=r_{g/\mathbb{P}}\,w_{qg}
\int_{x}^{1}\! dz\, z^{\alpha_{\mathbb P}(0)-1} &&\nonumber \\
\times \;P_{qg}(z)\,(1-x/z)^{\beta_{g/h}}\,(1+x/z)^{b_h},\label{gamma-q}&&
\end{eqnarray}
where $P_{qg}$ is the usual Altarelli-Parisi
splitting kernel for three active flavors and
the parameter $r_{g/\mathbb{P}}$ characterizes  gluon density in the 
soft Pomeron in the low $x$ limit,
 when probed at the virtuality scale  $Q_0^2$. We use $\beta_{g/p}=4$
 and $\beta_{g/\pi}=\beta_{g/K}=2$, while the constants $b_h$ are
 fixed requiring momentum conservation for parton distribution functions (PDFs).

 By construction, the eikonal $\chi_{Ih}^{\mathbb{P}_{{\rm soft}}}$ is related
 to the generalized parton distribution (GPD) $G_{I/h}$
  at the virtuality scale 
  $Q_0^2$ \cite{ost06,ost16}:
 \begin{equation}
x\,G_{I/h}(x,b,Q_0^2)= \chi_{Ih}^{\mathbb{P}_{{\rm soft}}}(s_{0}/x,b)\,.
\label{eq:gpd}
\end{equation}

It is further noteworthy that the above-discussed ``semihard Pomeron'' approach
largely resembles the ``heterotic Pomeron'' concept proposed in \cite{lev93}
(see also \cite{bon04}).
In particular, the relatively large slope $\alpha_{\mathbb P}'$ of the soft Pomeron
gives rise to a rather fast transverse expansion of  parton ``clouds''.
 On the other hand, the perturbative ($|q^2|>Q_0^2$) parton evolution,
 being characterized by small transverse displacements ($\lesssim 1/ Q_0$),
 leads to a quick rise of  parton density.

\section{Treatment of ``color fluctuations'' \label{gw.sec}}
One of the drawbacks of the scheme discussed so far is that it offers no room
for inelastic diffraction. To overcome that, one has to  account for
transitions of  interacting hadrons into various excited states, after each
``elementary'' rescattering process (Pomeron exchange) \cite{gri69}.
This can be conveniently done following the Good-Walker (GW) 
approach \cite{goo60} (see, e.g., \cite{kho21} for a recent discussion):
assuming both the original hadron $h$ and its excited states $h^*$ to be represented by a superposition of eigenstates of the scattering matrix:
\begin{eqnarray}
|h\rangle=\sum_{i}\sqrt{C^{(i)}_{h}}\,|i\rangle \label{eq:gw-h}\\
|h^*\rangle=\sum_{i}\sqrt{C^{(i)}_{h^*}}\,|i\rangle , \label{eq:gw-h*}
\end{eqnarray}
with $C^{(i)}_{h}$, $C^{(i)}_{h^*}$ being the corresponding partial weights
($\sum_{i}C^{(i)}_{h}=\sum_{i}C^{(i)}_{h^*}=1$).

In such an approach, one arrives to a trivial generalization of 
 Eqs.\ (\ref{eq:sig-eik-tot}--\ref{eq:sig(n)}),
averaging over  different combinations of GW Fock states:
\begin{eqnarray}
\sigma^{\rm tot}_{hp}(s) 
=  2\int \! d^2b \sum_{i,j}C^{(i)}_{h}C^{(j)}_{p}  && \nonumber \\
\times \;
\left[1-e^{-\chi_{hp(ij)}^{\mathbb P}(s,b)}\right]
\label{eq:sig-eik-tot-gw} &&\\
\sigma^{\rm el}_{hp}(s) =   \int \! d^2b && \nonumber \\
\times \left[\sum_{i,j}C^{(i)}_{h}C^{(j)}_{p}\,
(1-e^{-\chi_{hp(ij)}^{\mathbb P}(s,b)})\right]^2
\label{eq:sig-eik-el-gw}  && \\
\sigma^{(n)}_{hp}(s) = \int \! d^2b  \sum_{i,j}C^{(i)}_{h}C^{(j)}_{p}\, && \nonumber \\
\times \; \frac{\left[2\chi_{hp(ij)}^{\mathbb P}(s,b)\right]^n}{n!}\,
  \exp (-2\chi_{hp(ij)}^{\mathbb P}(s,b))\,.
   \label{eq:sig(n)-gw}  &&
\end{eqnarray}
Here the eikonals 
 $\chi^{\mathbb P}_{hp(ij)}=\chi^{{\mathbb P}_{\rm soft}}_{hp(ij)}
 +\chi^{{\mathbb P}_{\rm sh}}_{hp(ij)}$ correspond to exchanges of both
 soft and semihard Pomerons between  GW states $|i\rangle$ and $|j\rangle$ 
 of the projectile and the target, respectively.

In turn, the inelastic cross section is now split into the ``absorptive''
and diffractive parts:
\begin{equation}
\sigma^{\rm inel}_{hp}(s)\equiv \sigma^{\rm tot}_{hp}(s) -\sigma^{\rm el}_{hp}(s) =\sigma^{\rm abs}_{hp}(s) +\sigma^{\rm diffr}_{hp}(s)\,,
\end{equation}
with the former corresponding to any number ($n\geq 1$) of cut Pomeron exchanges,
\begin{eqnarray}
\sigma^{\rm abs}_{hp}(s) =\sum_{n=1}^{\infty} \sigma^{(n)}_{hp}(s) &&\nonumber \\
=\, \int \! d^2b \sum_{i,j}C^{(i)}_{h}C^{(j)}_{p}\,
[1-e^{-2\chi_{hp(ij)}^{\mathbb P}(s,b)}]\,, &&
\label{eq:sig-eik-abs-gw} 
\end{eqnarray}
and the latter containing contributions of the projectile, target, and double
diffraction.\footnote{See, e.g., \cite{ost11} for the corresponding partial cross sections.}

The corresponding soft Pomeron exchange eikonal is defined as
\begin{eqnarray}
\chi_{hp(ij)}^{{\mathbb P}_{\rm soft}}(s,b)=\frac{\gamma_{h(i)}\gamma_{p(j)}
(s/s_0)^{\alpha_{\mathbb P}(0)-1}}{R_{h(i)}^2+R_{p(j)}^2
+\alpha_{\mathbb P}'(0)\ln (s/s_0)}\,
&& \nonumber \\
\times \, \exp \!\left[- \frac{b^2/4}{R_{h(i)}^2+R_{p(j)}^2
+\alpha_{\mathbb P}'(0)\ln (s/s_0)}\right]\!,
\label{eq:chi-pom-gw} &&
\end{eqnarray}
taking into consideration that different GW Fock
states are generally characterized by
different sizes and different couplings to the Pomeron. Moreover, it is quite
reasonable to assume that the coupling $\gamma_{h(i)}$ is approximately
proportional to the transverse area of the state \cite{kai82}:
 \begin{equation}
\gamma_{h(i)}=g_0\,R_{h(i)}^2,\label{eq:gamma-r} 
\end{equation}
using thus a universal parameter $g_0$.

The ``semihard Pomeron'' eikonal is generalized similarly:
\begin{eqnarray}
\chi_{hp(ij)}^{\mathbb{P}_{{\rm sh}}}(s,b)=\frac{1}{2}\sum_{I,J}
\int\! d^{2}b'\int\!\frac{dx^{+}}{x^{+}}\frac{dx^{-}}{x^{-}}  &&\nonumber \\
\times\;\chi_{Ih(i)}^{\mathbb{P}_{{\rm soft}}}(\frac{s_{0}}{x^{+}},b') \;
\chi_{Jp(j)}^{\mathbb{P}_{{\rm soft}}}(\frac{s_{0}}{x^{-}},|\vec{b}-\vec{b}'|)
  &&\nonumber \\ \times\;
\sigma_{IJ}^{{\rm QCD}}(x^{+}\, x^{-}\,s,Q_{0}^{2},Q_{0}^{2})\,,
\label{chi-sh-gw}&&
\end{eqnarray}
where the eikonal $\chi_{Ih(i)}^{\mathbb{P}_{{\rm soft}}}$ corresponds
  to a soft Pomeron exchange between parton $I$ and   GW state $|i\rangle$
  of   hadron $h$:
\begin{eqnarray}
\chi_{Ih(i)}^{\mathbb{P}_{{\rm soft}}}(\hat{s},b) =\frac{\gamma_{h(i)}
V^{\mathbb P}_{I/h(i)}(s_{0}/\hat{s})
(\hat{s}/s_{0})^{\alpha_{\mathbb P}(0)-1}}
{R_{h(i)}^{2}+\alpha_{\mathbb{P}}'(0)\,\ln(\hat{s}/s_{0})}&&\nonumber \\
\times \;\exp\!\left[-\frac{b^{2}/4}{R_{h(i)}^{2}
+\alpha_{\mathbb{P}}'(0)\,\ln(\hat{s}/s_{0})}\right]\!,\label{chi_aI-gw}&&
\end{eqnarray}
with $V^{\mathbb P}_{I/h(i)}$ being defined by 
Eqs.\ (\ref{gamma-g}-\ref{gamma-q}), under the replacement
$b_h  \rightarrow b_{h(i)}$.
Here we have an important feature: assuming a universal  gluon density
 $r_{g/\mathbb{P}}$ in the soft Pomeron in the low $x$ limit, at the
  scale  $Q_0^2$,  the momentum conservation for PDFs
   for {\em individual} Fock states,
 \begin{eqnarray}
 x\,f_{I/h(i)}(x,Q_0^2)=\int \!d^2b\; x\,G_{I/h(i)}(x,b,Q_0^2)&&\nonumber \\
 =\int \!d^2b\;   \chi_{Ih(i)}^{\mathbb{P}_{{\rm soft}}}(s_{0}/x,b)\,,
\label{eq:gpd-gw} &&
\end{eqnarray}
gives rise to different $b_{h(i)}$ for different states.
Consequently, partial PDFs $f_{I/h(i)}$ vary considerably from one Fock state 
to another  \cite{fra08}: with smaller Fock states being characterized by 
  smaller (integrated) parton
densities in the low $x$ limit but having harder PDF shapes.

It is noteworthy, however, that the above-discussed approach is applicable,
 strictly speaking, to the treatment of low mass diffraction only. Indeed,
 considering hadron $h$ transitions into multiparticle states $h^*$ of arbitrary
 mass, one can no longer apply the decomposition,
  Eqs.\ (\ref{eq:gw-h}-\ref{eq:gw-h*}), based on a 
 finite number of GW states:\footnote{In principle, one may consider a continuum
 of intermediate multiparticle states, postulating some energy and mass
 dependence for the decomposition in Eqs.\ (\ref{eq:gw-h}-\ref{eq:gw-h*}).}
 with increasing energy, larger and larger excited states will play an
 important role.

 In the following, we consider equal probabilities for  different GW
 states, $C^{(i)}_{h} \equiv 1/N_{\rm GW}$, using $N_{\rm GW}=3$ for any 
 hadron\footnote{We verified explicitly that using a larger number of GW states
 does not modify our results significantly.} and choosing a 
 loguniform distribution for $R^2_{h(i)}$:
\begin{eqnarray}
R^2_{h(i)}=R^2_{h(1)}\,d_h^{\frac{i-1}{N_{\rm GW}-1}}\label{eq:R_i}.
\end{eqnarray}

\section{Enhanced Pomeron diagrams \label{enhanced.sec}}
\begin{figure*}[t]
\begin{centering}
\includegraphics[width=\textwidth,height=2.5cm]{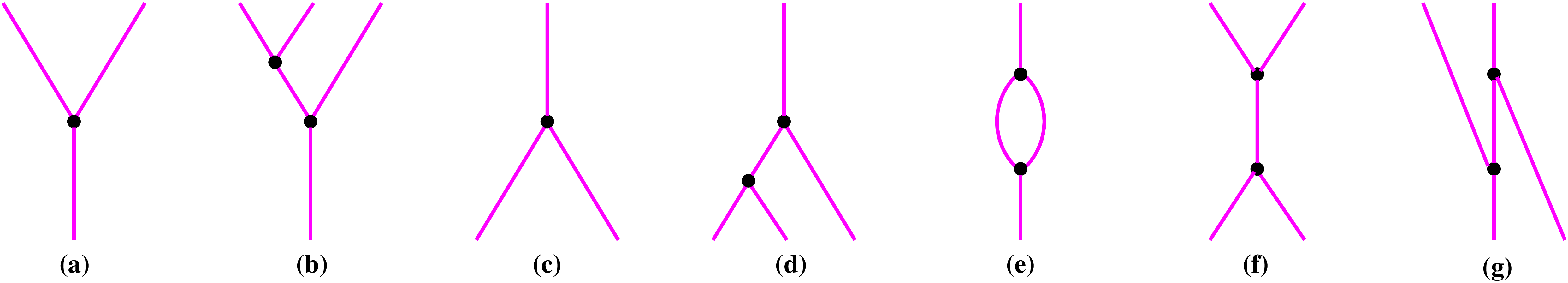}
\par\end{centering}
\caption{Simplest examples of  enhanced Pomeron diagrams.
\label{Fig:low-enh}}
\end{figure*}

The major feature inherited from the previous model version,  QGSJET-II \cite{ost06,ost11}, is   a treatment
of nonlinear interaction effects, based on all-order resummation of the
underlying Pomeron-Pomeron interaction diagrams
 \cite{ost06a,ost08,ost10}. The simplest examples of such, so-called enhanced,
 graphs are shown in Fig.\ \ref{Fig:low-enh}, 
 which correspond to rescattering of intermediate
 partons in the ``elementary'' parton cascades off the projectile
 (a, b, f, g) and  the target (c, d, f, g)
 hadrons or off each other (e). 
 Taking such diagrams into consideration amounts
 to replace the eikonal $\chi^{\mathbb P}_{hp(ij)}$ in 
 Eqs.\ (\ref{eq:sig-eik-tot-gw}--\ref{eq:sig(n)-gw})
 by  $\chi^{\rm scr}_{hp(ij)}=\chi^{\mathbb P}_{hp(ij)} +
 \chi^{\rm enh}_{hp(ij)}$,
  where   $\chi^{\rm enh}_{hp(ij)}$ corresponds to the
 summary contribution of all significant irreducible enhanced Pomeron graphs
 exchanged between   GW Fock states $|i\rangle$ and $|j\rangle$
  of the projectile and the target, respectively. 
 
 Using eikonal  vertices for the transition of $m$ into
$n$ Pomerons ($m+n\geq3$) \cite{kai86},
\begin{equation}
G^{(m,n)}=G\,\gamma_{\mathbb{P}}^{m+n},\label{eq:g_mn}
\end{equation}
with the constant $G$ being related to the triple-Pomeron coupling $r_{3\mathbb{P}}$
as $G=r_{3\mathbb{P}}/(4\pi\gamma_{\mathbb{P}}^{3})$, and neglecting the small
slope of the triple-Pomeron vertex $R^2_{3\mathbb{P}}$,
one was able to obtain $\chi^{\rm enh}_{hp(ij)}$ in a relatively
 compact form \cite{ost08,ost10,ost11}: 
\begin{eqnarray}
\chi_{hp(ij)}^{{\rm enh}}(s,b)=G \! \int_{\xi}^{Y-\xi}\! dy'\!\int\!\! d^{2}b'
\nonumber \\
\times \,\{ [(1-e^{-\chi_{h(i)|p(j)}^{{\rm net}}})
 \,(1-e^{-\chi_{p(j)|h(i)}^{{\rm net}}})
\nonumber \\-\,\chi_{h(i)|p(j)}^{{\rm net}}\,\chi_{p(j)|h(i)}^{{\rm net}}]
-[\chi_{h(i)|p(j)}^{{\rm net}}-\chi_{h(i)}^{{\rm loop}}]\nonumber \\
\times \,[(1-e^{-\chi_{p(j)|h(i)}^{{\rm net}}})\,e^{-\chi_{h(i)|p(j)}^{{\rm net}}}
-\chi_{p(j)|h(i)}^{{\rm net}}]\nonumber \\
+\,\chi_{p(j)}^{\mathbb{P}}(y',b')\,[\chi_{h(i)}^{{\rm loop}}
-\chi_{h(i)}^{{\rm loop(1)}}]\} .\label{eq:chi-enh}
\end{eqnarray}
Here $Y=\ln (s/s_0)$,  $\xi$ is the
minimal rapidity ``size'' of the Pomeron, and
 the omitted arguments of the eikonals read $\chi_{h(i)|p(j)}^{{\rm net}}=\chi_{h(i)|p(j)}^{{\rm net}}(Y-y',\vec{b}-\vec{b}'|Y,\vec{b})$,
$\chi_{p(j)|h(i)}^{{\rm net}}=\chi_{p(j)|h(i)}^{{\rm net}}(y',\vec{b}'|Y,\vec{b})$, 
$\chi_{h(i)}^{{\rm loop}}=\chi_{h(i)}^{{\rm loop}}(Y-y',|\vec{b}-\vec{b}'|)$,
$\chi_{h(i)}^{{\rm loop(1)}}=\chi_{h(i)}^{{\rm loop(1)}}(Y-y',|\vec{b}-\vec{b}'|)$.

The ``net-fan'' contributions $\chi_{h(i)|p(j)}^{{\rm net}}$ are defined
by a recursive equation:
\begin{eqnarray}
\chi_{h(i)|p(j)}^{{\rm net}}(y',\vec{b}'|Y,\vec{b})=
\chi_{h(i)}^{{\rm loop}}(y',b') +G \!\int\!\! d^{2}b''\nonumber \\
\times \int_{\xi}^{y'-\xi}\! dy''\,
(1-e^{-\chi^{{\rm loop}}(y'-y'',|\vec{b}'-\vec{b}''|)})
\nonumber \\
\times \left[(1-
e^{-\chi_{h(i)|p(j)}^{{\rm net}}(y'',\vec{b}''|Y,\vec{b})})\right.
\nonumber \\
\times\, \exp(-\chi_{p(j)|h(i)}^{{\rm net}}(Y-y'',\vec{b}-\vec{b}''|Y,\vec{b}))
\nonumber \\
 -\left. \chi_{h(i)|p(j)}^{{\rm net}}(y'',\vec{b}''|Y,\vec{b})\right] \!,
 \label{net-fan}
 \end{eqnarray}
where  $\chi_{h(i)}^{{\rm loop}}$ and  $\chi_{h(i)}^{{\rm loop(1)}}$
correspond to general irreducible two-point sequences of Pomerons and 
Pomeron loops, exchanged between   GW state $|i\rangle$ of
 hadron $h$ and a multi-Pomeron vertex, while
$\chi^{{\rm loop}}$ and $\chi^{{\rm loop(1)}}$ are contributions of such sequences
exchanged between two multi-Pomeron  vertices (see \cite{ost11} for more details).
Neglecting Pomeron loop insertions,
  $\chi_{h(i)}^{{\rm loop}}$ and  $\chi_{h(i)}^{{\rm loop(1)}}$ reduce to the
 eikonal 
 $\chi_{h(i)}^{\mathbb{P}}=\chi^{{\mathbb P}_{\rm soft}}_{h(i)}
 +\chi^{{\mathbb P}_{\rm sh}}_{h(i)}$ describing an exchange of a
 soft or semihard Pomeron  between the hadron $h$ represented by its GW state
 $|i\rangle$  and a multi-Pomeron  vertex. Here
  $\chi^{{\mathbb P}_{\rm soft}}_{h(i)}(y,b)$ is obtained from 
  Eq.\ (\ref{eq:chi-pom-gw}), for $s=s_0\,e^y$, replacing the vertex factor
   $\gamma_{p(j)}$ by $\gamma_{\mathbb{P}}$ and $R_{p(j)}^2$ by 
   $R_{3\mathbb{P}}^2\simeq 0$:
\begin{eqnarray}
\chi^{{\mathbb P}_{\rm soft}}_{h(i)}(y,b)=\frac{\gamma_{h(i)}\gamma_{\mathbb{P}}\,
e^{(\alpha_{\mathbb P}(0)-1)y}}{R_{h(i)}^2+\alpha_{\mathbb P}'(0)\,y}
e^{\frac{-b^2/4}{R_{h(i)}^2+\alpha_{\mathbb P}'(0)\,y}} .
\label{eq:chi-pom-pom} 
\end{eqnarray}
In turn, $\chi^{{\mathbb P}_{\rm sh}}_{h(i)}(y,b)$ is obtained from
 Eq.\ (\ref{chi-sh-gw}), for $s=s_0\,e^y$,
  replacing $\chi_{Jp(j)}^{\mathbb{P}_{{\rm soft}}}$ by
 $\chi_{J\mathbb{P}}^{\mathbb{P}_{{\rm soft}}}$, with
\begin{eqnarray}
\chi_{J\mathbb{P}}^{\mathbb{P}_{{\rm soft}}}(\hat{s},b) =\frac{\gamma_{\mathbb{P}}\,
V^{\mathbb P}_{J}(s_{0}/\hat{s})\,
(\hat{s}/s_{0})^{\alpha_{\mathbb P}(0)-1}}
{\alpha_{\mathbb{P}}'(0)\ln(\hat{s}/s_{0})}&&\nonumber \\
\times \;\exp\!\left[-\frac{b^{2}}{4\alpha_{\mathbb{P}}'(0)\ln(\hat{s}/s_{0})}\right],
\label{eq:chi_IP}&&
\end{eqnarray}
where  $V^{\mathbb P}_{J}$ is defined by 
Eqs.\ (\ref{gamma-g}-\ref{gamma-q}), for
$\beta_{g/h}= \beta_{g/\pi}$.

In contrast to \cite{ost06,ost11}, where $\gamma_{\mathbb{P}}$ was treated as an
adjustable parameter, here we make a specific choice:
\begin{equation}
\gamma_{\mathbb{P}}=r_{3\mathbb{P}}/(\alpha_{\mathbb P}(0)-1).
\label{eq:gamma-pom-crit} 
\end{equation}
Taking into account the ``renormalization'' of the soft Pomeron
in the ``dense'' (high $s$, small $b$) limit \cite{kai86}, this pushes it into
the ``critical'' regime in such a limit (see \cite{rys11} for a recent
 discussion):  with the renormalized Pomeron intercept
\begin{equation}
\alpha_{\mathbb P}^{\rm (ren)}(0)=
\alpha_{\mathbb P}(0)-r_{3\mathbb{P}}/\gamma_{\mathbb{P}}=1\,.
\label{eq:gamma-pom} 
\end{equation}
In turn, this leads to a ``saturation'' of both partial gluon and 
sea (anti)quark GPDs  $G_{I/h(i)}^{\rm scr}$
and of the corresponding total GPDs,
\begin{equation}
G_{I/h}^{\rm scr}(x,b,Q_0^2)=\sum_i C^{(i)}_{h}\,G_{I/h(i)}^{\rm scr}(x,b,Q_0^2)\,,
\label{eq:gpd-h} 
\end{equation}
 at the virtuality scale   $Q_0^2$, in the low $x$ 
 {\em and small $b$} limit, for any hadron $h$ \cite{ost19}. 

Here, taking into account absorptive corrections due to enhanced Pomeron
diagrams, $G_{I/h(i)}^{\rm scr}$ are defined 
as\footnote{In  \cite{ost06,ost16},
a simplified expression for $G_{I/h(i)}^{\rm scr}$ had been provided,
neglecting Pomeron loop contributions.} \cite{ost06,ost16}
\begin{eqnarray}
x\,G_{I/h(i)}^{\rm scr}(x,b,Q_0^2)
\nonumber \\
=\; \chi_{Ih(i)}^{\mathbb{P}_{{\rm soft}}}(s_{0}/x,b)
+G\! \int\!\! d^{2}b'
\nonumber \\
\times \;\int_{\xi}^{-\ln x}\!dy'
 \left\{\chi_{I\mathbb{P}}^{\mathbb{P}_{{\rm soft}}}(s_{0}e^{-y'}/x,|\vec{b}-\vec{b}'|)
\right.\nonumber \\
\times  \left[\chi_{h(i)}^{{\rm loop}}(y',b')-\chi_{h(i)}^{{\rm loop(1)}}(y',b')\right]
\nonumber \\
 +\, \chi_{I}^{\rm loop}(-\ln x-y',|\vec{b}-\vec{b}'|)
\nonumber \\
\times \left.\left[1-e^{-\chi_{h(i)}^{{\rm fan}}(y',b')}-\chi_{h(i)}^{{\rm fan}}(y',b')\right]\right\} \!.
\label{eq:GPD-Q0}
\end{eqnarray}
Here $\chi_{h(i)}^{{\rm fan}}$ is a solution of the ``fan'' diagram
equation [c.f.\ Eq.\ (\ref{net-fan})]:
\begin{eqnarray}
\chi_{h(i)}^{{\rm fan}}(y',b')=\chi_{h(i)}^{{\rm loop}}(y',b')+G\int\! d^{2}b''
\nonumber \\
\times \;\int_{\xi}^{y'-\xi}\! dy''
 \left[1-e^{-\chi^{\rm loop}(y'-y'',|\vec{b}'-\vec{b}''|)}\right]\nonumber \\
\times\left[1-e^{-\chi_{h(i)}^{{\rm fan}}(y'',b'')}-\chi_{h(i)}^{{\rm fan}}(y'',b'')\right],
\label{eq:fan}
\end{eqnarray}
while $\chi_{I}^{\rm loop}$ is defined as
\begin{eqnarray}
\chi_{I}^{\rm loop}(y',b')=\chi_{I\mathbb{P}}^{\mathbb{P}_{{\rm soft}}}(s_{0}e^{y'},b')+G\int\! d^{2}b''
\nonumber \\
\times \;\int_{\xi}^{y'-\xi}\! dy''\;
 \chi_{I\mathbb{P}}^{\mathbb{P}_{{\rm soft}}}(s_{0}e^{y'-y''},|\vec{b}'-\vec{b}''|)
\nonumber \\
\times \left[1-e^{-\chi^{\rm loop}(y'',b'')}
-\chi^{\rm loop(1)}(y'',b'')\right].
\label{eq:Iloop}
\end{eqnarray}

Applying the AGK cutting rules \cite{agk74}, one was able to obtain the complete set
of unitarity cuts of elastic scattering diagrams for hadron-hadron collisions,
corresponding to the above-discussed resummation scheme, explicitly verifying
the $s$-channel unitarity of the approach, and to derive 
positive-definite partial cross sections
for all the various configurations of final states, 
including diffractive ones \cite{ost08,ost10}.
In turn, those allowed one to develop a MC procedure for generating such
configurations both for hadron-proton and for hadron-nucleus (nucleus-nucleus)
scattering events \cite{ost11}.

As discussed in \cite{ost06,ost16}, an important feature of the described approach is a consistency with the
collinear factorization of pQCD: by virtue of the AGK cancellations \cite{agk74},
 the inclusive parton jet production cross section is defined by the usual
factorization ansatz: 
\begin{eqnarray}
\frac{d\sigma_{hp}^{\rm jet}(s,p_{{\rm t}})}{dp_{{\rm t}}^{2}}
=K \!\int\! dx^{+}dx^{-}\sum_{I,J}f_{I/h}^{\rm scr}(x^{+},\mu_{{\rm F}}^{2}) \nonumber \\
\times \,f_{J/p}^{\rm scr}(x^{-},\mu_{{\rm F}}^{2})\,
\frac{2\,d\sigma_{IJ}^{2\rightarrow2}(x^{+}x^{-}s,p_{t}^{2},\mu_{\rm R})}
{dp_{{\rm t}}^{2}}\,.\label{eq:sig-2jet}
\end{eqnarray}
Here the PDFs $f_{I/h}^{\rm scr}$, with the absorptive corrections taken into account, are expressed via the partial GPDs   $G_{I/h(i)}^{\rm scr}$,
 Eq.\ (\ref{eq:GPD-Q0}), evolving the latter from $Q_0^2$ to $\mu_{{\rm F}}^{2}$:
\begin{eqnarray}
f_{I/h}^{\rm scr}(x,\mu_{{\rm F}}^{2})=\sum_i C^{(i)}_{h} \int\! d^{2}b
 \sum _{I'}\int_x^1 \!\frac{dz}{z}
\nonumber \\
\times \,E_{I'I}^{{\rm QCD}}(z,Q_{0}^{2},\mu_{{\rm F}}^{2})\,
G^{\rm scr}_{I'/h(i)}(x/z,b,Q_0^2)\,.
\label{eq:pdfs-scr} 
\end{eqnarray}

\section{Dynamical power corrections to hard scattering\label{ht.sec}}
As mentioned in Section \ref{enhanced.sec}, an important feature of the above-discussed approach is the consistency with the
collinear factorization of pQCD: the inclusive jet cross section is defined
by Eq.\ (\ref{eq:sig-2jet}). However, this creates an unpleasant sensitivity
of the model predictions 
to the choice of the ``infrared'' cutoff $Q_0^2$: since 
$d\sigma_{hp}^{\rm jet}(s,p_{{\rm t}})/dp_{{\rm t}}^{2}$ explodes in the small
$p_t$ limit. For example, in case of the QGSJET-II model, a reasonable consistency with collider measurements is reached for a rather high value
of that cutoff, $Q_0^2=3$ GeV$^2$ \cite{ost11}. On the other hand, one may expect
the pQCD approach to remain applicable for parton virtualities as
small as $\sim 1$   GeV$^2$. It is thus natural to ask ourselves whether there 
exists a perturbative mechanism capable of damping the jet production at small
$p_t$.

To address this question, it is useful to remind oneself that the collinear
 factorization of pQCD is established at the leading twist level \cite{col88,col89}, i.e.,
 neglecting the so-called higher twist (HT) corrections suppressed by powers of
 the relevant hard scale. One may thus  expect
 that those are such power corrections which should provide the desirable
 suppression of low $p_t$ jet production.
 
 Unfortunately it is hardly possible at the present stage to treat
HT effects in  hadronic collisions in a  systematic way, especially,
 regarding their potential
 implementation in MC event generators: in particular, since
this involves a significant number of unknown
 multiparton correlators and  the corresponding HT
 contributions are not generally
 positive-definite, excluding thereby a probabilistic interpretation \cite{jaf81,ell82,ell83}. 
 Therefore, we adopt here a phenomenological approach \cite{ost19,ost19a}, 
 concentrating on a 
 particular class of   dynamical power corrections
  to  parton scattering processes, corresponding to 
coherent multiple rescattering of $s$-channel partons on virtual ``soft''
[characterized by  small light cone (LC)
 momentum fractions, $x_g\sim 0$] gluon pairs  \cite{qiu04,qiu04a,qiu06}.
Such contributions have been shown  to provide dominant nuclear 
size-enhanced power corrections to the low $x$ and low $Q^2$ behavior of 
structure functions (SFs) in deep inelastic scattering (DIS) 
on nuclear targets \cite{qiu04,qiu04a}
  and to the suppression of jet  $p_t$-spectra in high energy proton 
  scattering on heavy nuclei, for moderately small  $p_t$ \cite{qiu06}.

\subsection{Resummed $A$-enhanced power corrections \cite{qiu04,qiu04a,qiu06}\label{ht-a.sec}}
\begin{figure*}[t]
\centering
\includegraphics[height=3.5cm,width=0.4\textwidth]{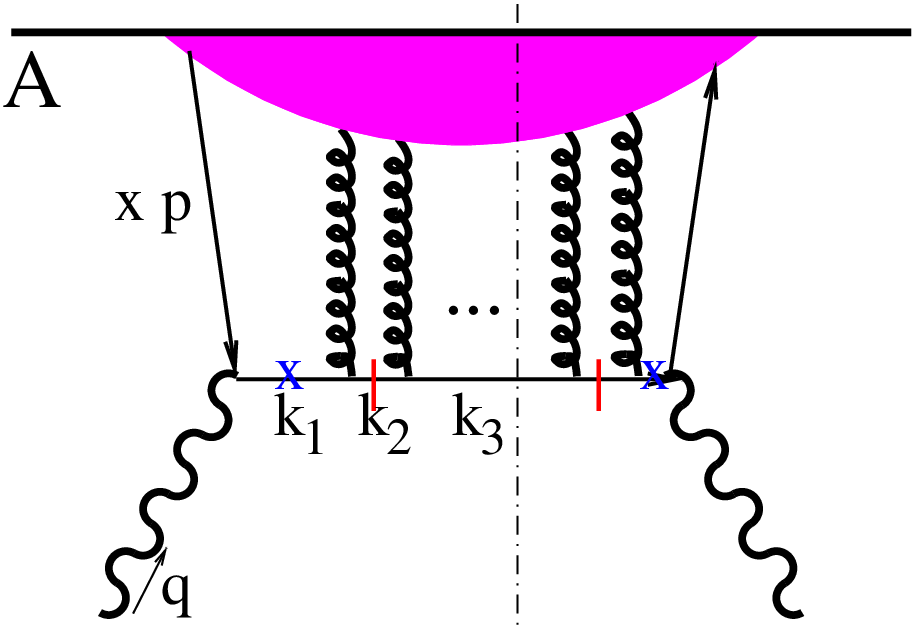}\hfill
\includegraphics[height=3.5cm,width=0.4\textwidth]{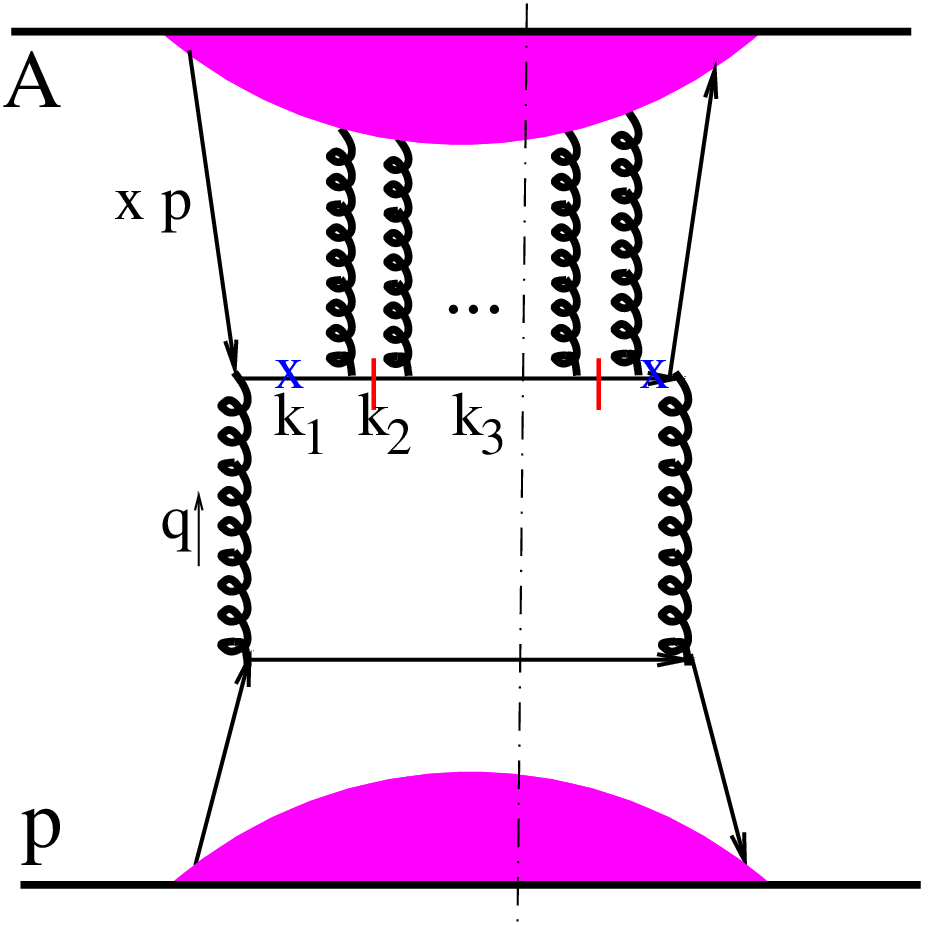}
\caption{Example diagrams for dominant $A$-enhanced power corrections to nuclear DIS (left)
and to jet production in proton-nucleus scattering (right). Propagators marked by crosses
and vertical dashes correspond to pole and contact terms, respectively.}
\label{fig:A-ht}       
\end{figure*}

 Regarding  HT corrections to nuclear SFs, the dominant
 $A$-enhanced contributions, in the small Bjorken $x$ limit,
  were shown to come from diagrams of the kind depicted in 
 Fig.\ \ref{fig:A-ht} (left), 
corresponding to a rescattering of the struck quark on soft, pairwise singlet, 
virtual gluon pairs \cite{qiu04,qiu04a}.
  The perturbative hard part of the graph is
 characterized by a specific structure of the $s$-channel quark propagators:
 with the gluon fields in a pair being
separated by the so-called ``contact'' term,  implying
no propagation along the LC coordinate \cite{qiu90}; propagators which 
separate such gluon pairs  from the quark field and from each other are 
represented, on the other hand, by  pole terms corresponding to
 a propagation over considerable LC distances.
It is such a prescription which gives rise to the nuclear enhancement:
in the low $x$ limit, the struck quark from one nucleon in the nucleus
 propagates over large distances $\propto 1/(x\,p^+)$, 
 $p^+$ being the LC-plus (LC$^+$) momentum of the nucleon, scattering coherently
 on many correlated soft gluon pairs from other
  nucleons  \cite{qiu04,qiu04a}. Upon the full all-order resummation of such 
  contributions (in the LC $A^+=0$ gauge), one obtained the HT correction to the 
  nuclear transverse structure  function $F_{\rm T}^{(A)}$ (c.f.\ Eq.\ (8) in \cite{qiu04}):
\begin{eqnarray}
\delta _{HT}F_{\rm T}^{(A)}(x,Q^2) \nonumber \\
=\; \sum _q \frac{e_q^2}{2}\sum _{n\geq 1}  \frac{1}{n!}
\left[\frac{4\pi^2\alpha_sx}{3Q^2}\right]^n
 \frac{d^n}{d^n\!x} T^{(n)}_{q\underset{n}{\underbrace{gg...}}}(x)\,,
    \label{eq:HT-F2A}
   \end{eqnarray}
where  $e_q$ is the fractional charge of (anti)quark  $q$  and
the multiparton correlators  $T^{(n)}_{qgg...}$ are defined as
\begin{eqnarray}
T^{(n)}_{q\underset{n}{\underbrace{gg...}}}(x) =
  \int \! \frac{dy^-}{4\pi}\; e^{ip^+x\,y^-}  
   \langle A|\bar \psi(0)\,\gamma^+\,
\nonumber \\
 \times \,
\prod _{i=1}^n \left[ \int \! p^+dy^-_{g_i}\,\Theta(y^-_{g_i})\, \hat F^{2}(y^-_{g_i}) \right] 
\psi(y^-)|A\rangle ,
\label{eq:T-ht-q}       
 \end{eqnarray}
 with
\begin{eqnarray}
 \hat F^{2}(y^-_{g})=\int \! \frac{d\tilde y^-_{g}}{2\pi\,p^+}\;
F^{+\alpha}(y^-_{g})\,F^+_{\alpha}(\tilde y^-_{g})
\nonumber \\
 \times \; \Theta(y^-_{g}-\tilde y^-_{g})\, .
\label{eq:fg2-ht}       
 \end{eqnarray}
Here $y^-$, $y^-_{g_i}$ are LC-minus (LC$^-$) 
  coordinates of the fields and $F^+_{\alpha}$ is
 the projection of the gluon field tensor on the LC$^+$ direction.

Likewise, in case of nucleus-proton scattering, the dominant
 $A$-enhanced HT corrections to jet production cross section
  arise from the diagrams of 
 Fig.\ \ref{fig:A-ht} (right), corresponding to a rescattering of the 
 nuclear parton  (quark or gluon), participating in the hard scattering,
 on such soft gluon pairs, with the same structure of $s$-channel propagators
 along the struck parton line\footnote{Alternative configurations of the hard
 scattering part of the graph contain contact terms leading to a loss of the
 $A$-enhancement \cite{qiu06}.} \cite{qiu06}. Resumming such contributions 
 to all orders, one obtained \cite{qiu06}:
\begin{eqnarray}
\delta _{HT} \frac{d\sigma_{pA}^{\rm jet}(s,p_{{\rm t}})}{dp_{{\rm t}}^{2}}
=\int\! dx^{+}dx^{-} \nonumber \\
\times \,\sum_{I,J}f_{J/p}(x^{-},\mu_{{\rm F}}^{2})
\sum _{n\geq 1}  \frac{1}{n!}
\left[\frac{-C_I\pi^2\alpha_sx^+}{\hat t}\right]^n\nonumber \\
\times \, \frac{d^n}{d^n\!x} \left[
T^{(n)}_{I\underset{n}{\underbrace{gg...}}}(x^+)\,
\frac{2\,d\sigma_{IJ}^{2\rightarrow2}}
{dp_{{\rm t}}^{2}}\right],
\label{eq:sig-2jet-ht}
 \end{eqnarray}
 with  $C_{q(\bar q)}=C_F=4/3$, $C_g=C_A=3$, and $\hat t=q^2$ being the momentum
 transfer squared for parton-parton scattering.
 In addition to  $T^{(n)}_{qgg...}$ defined by Eqs.\ (\ref{eq:T-ht-q}-\ref{eq:fg2-ht}),
  Eq.\  (\ref{eq:sig-2jet-ht}) involves multigluon correlators  $T^{(n)}_{ggg...}$:
\begin{eqnarray}
T^{(n)}_{g\underset{n}{\underbrace{gg...}}}(x) =
  \int \! \frac{dy^-}{2\pi x p^+}\; e^{ip^+x\,y^-}  
   \langle A|F^{+\beta}(0) 
\nonumber \\ \times \,
\prod _{i=1}^n \left[ \int \! p^+dy^-_{g_i}\Theta(y^-_{g_i}) 
\hat F^{2}(y^-_{g_i}) \right]  F^+_{\beta}(y^-)|A\rangle .
\label{eq:T-ht-g}       
 \end{eqnarray}

Further, one proposed a model for the multiparton correlators
 $T^{(n)}_{Igg...}$, $I=q,g$: assuming that the dominant contributions to 
 Eqs.\ (\ref{eq:HT-F2A}) and (\ref{eq:sig-2jet-ht}) arise
  from a rescattering of the struck
 nuclear parton on gluon pairs belonging to different nucleons and
  performing the corresponding factorization \cite{qiu04}:
 \begin{equation}
 \langle A|\hat O_0\prod _{i=1}^n \hat O_i|A\rangle \propto \langle p|\hat O_0|p\rangle 
  \prod _{i=1}^n\langle p|\hat O_i|p\rangle .
\label{eq:A-fact}       
 \end{equation}
 This allowed one to obtain closed compact results for 
  $F_{\rm T}^{(A)}$  and $d\sigma^{\rm jet}_{pA}/dp_t^2$,
  the power corrections being accounted for:
\begin{eqnarray}
F_{\rm T}^{(A)}(x,Q^2) \simeq  A \sum _q \frac{e_q^2}{2} \nonumber \\
\times \; f_{q/p}(x+xC_q\xi^2(A^{1/3}-1)/Q^2,Q^2)
    \label{eq:HT-FT-A} \\
\frac{d\sigma_{pA}^{\rm jet}(s,p_{{\rm t}})}{dp_{{\rm t}}^{2}}
\simeq A\int\! dx^{+}dx^{-} \nonumber \\
\times \, \sum_{I,J} f_{I/p}(x^{+}-\frac{x^{+}C_I\xi^2(A^{1/3}-1)}{\hat t},\mu_{{\rm F}}^{2})
 \nonumber \\
\times \;f_{J/p}(x^{-},\mu_{{\rm F}}^{2})\,
 \frac{2\,d\sigma_{IJ}^{2\rightarrow2}}{dp_{{\rm t}}^{2}}\,,
\label{eq:sig-2jet-ht-A}
   \end{eqnarray}
where $\xi^2 \propto \langle p|\hat F^{2}|p\rangle \propto \lim _{x\rightarrow 0}x\,f_g/p$ defines the characteristic scale of the HT corrections.

\subsection{Phenomenological implementation in QGSJET-III\label{ht-q3.sec}}
Since we are going to extrapolate the treatment of \cite{qiu06} to the case
of hadron-proton scattering, we need a different approach for modeling the
corresponding multiparton correlators.
 Starting with the quark-gluon
correlator   $T^{(1)}_{qg}$, a closer look at Eq.\ (\ref{eq:T-ht-q}) reveals that
it formally coincides, up to a factor, with the quark-gluon  $^2$GPD
 $F^{(2)}_{qg}$ multiplied by the gluon
 LC momentum fraction $x_{g}$,  in the limit
  $x_{g}\rightarrow 0$ and for zero transverse separation between the two
 partons, $\vec \Delta=0$ (see, e.g.\ \cite{die12} for the corresponding definitions),
  and similarly for $T^{(1)}_{gg}$ in  Eq.\ (\ref{eq:T-ht-g}).
     This motivated us to employ a probabilistic
   treatment for $T^{(1)}_{qg}$ and  $T^{(1)}_{gg}$,    interpreting them as 
   $\propto \left.x_{g}\,F^{(2)}_{qg}\right|_{\vec \Delta=0}$ and 
    $\propto \left.x_{g}\,F^{(2)}_{gg}\right|_{\vec \Delta=0}$, respectively, 
   and to proceed in a similar way with all the other  correlators 
    $T^{(n)}_{Igg...}$     involving    larger numbers of soft gluons. 
   
    Here we have to
   make additional assumptions concerning the relevant virtuality scales
    and  gluon momentum fractions in the corresponding multiparton GPDs,
    e.g., for  $Q_q^{2},Q_g^{2}$, and   $x_{g}$  in 
   $F^{(2)}_{qg}(x,x_g,Q_q^{2},Q_g^{2},\vec{\Delta})$. While the
   natural choice for $Q_q^{2}$ is the factorization scale $\mu_{\rm F}^2$
   for the hard process, one usually considers soft gluons to be 
   purely nonperturbative  ones, with $Q_g^{2}\sim \Lambda_{\rm QCD}^2$
    (e.g.\ \cite{rau03}).
   Instead, we set $Q_g^{2}$ equal to our separation scale $Q_0^2$, in order
   to describe the GPDs by  soft Pomeron asymptotics, plus absorptive
   corrections, Eq.\ (\ref{eq:GPD-Q0}).

   Finally, assuming a finite  virtuality for the soft gluons implies that
   they have nonzero LC$^{\pm}$ momentum fractions $x_g^{\pm}$:
   \begin{equation}
    |q_g^2|\sim  x_g^+\,x^-_g\,s\,.
    \label{eq:soft-q2}
   \end{equation}
    In the factorization procedure which led to Eqs.\ (\ref{eq:HT-F2A})
    and (\ref{eq:sig-2jet-ht}),
    one neglected  LC$^-$  momentum components for projectile partons
    (similarly neglecting LC$^+$ momenta of target partons) and considered
 the limit   $x_{g}\rightarrow 0$ for the soft gluons involved in the process.
 Here, taking into account the  finite virtuality of such gluons, 
 Eq.\ (\ref{eq:soft-q2}), and the fact that these   gluons
   belong to the projectile proton (for the case of rescattering on 
   the projectile soft   gluons), 
 their LC$^-$ momentum fractions  should be much smaller than 
 the  LC$^-$ fraction of the
    target parton participating in the  hard process:
\begin{equation}
x^-_g \sim  \frac{|q_g^2|}{x_g^+\,s}\ll x^-\,.
\end{equation}
Since we expect a rather weak $x_g$-dependence for $x_{g}\,F^{(2)}_{qg}$ and
 $x_{g}\,F^{(2)}_{gg}$ in the small  $x_g$ limit at the low virtuality
 scale $Q_g^{2}=Q_0^2$, we  set
\begin{equation}
x_g=  \frac{Q_0^2}{x^-\,s}\,.
\end{equation}

Using these additional assumptions, we obtain the leading power correction
to the inclusive parton jet production cross section as 
[c.f.\ Eq.\ (\ref{eq:sig-2jet-ht})]
\begin{eqnarray}
\delta ^{(1)}_{HT}\frac{d\sigma_{hp}^{\rm jet}(s,p_{{\rm t}})}{dp_{{\rm t}}^{2}}
 \nonumber \\
=K\int\! dx^{+}dx^{-}\sum_{I,J}f_{J/p}^{\rm scr}(x^{-},\mu_{{\rm F}}^{2}) \nonumber \\
\times \;
\frac{K_{\rm HT}\,C_I\,\pi^2\,\alpha_{\rm s}(\mu_{\rm R}^2)\,x^+}{|\hat t|}\,
\frac{2\,d\sigma_{IJ}^{2\rightarrow2}(\hat s,p_{t}^{2},\mu_{\rm R})}
{dp_{{\rm t}}^{2}}  \nonumber \\
\times \left. x_g^+F^{(2)}_{Ig/h}(x^+,x_g^+,\mu_{\rm F}^2,Q_0^{2},
\vec \Delta =\vec 0)\right|_{x_g^+=\frac{Q_0^2}{x^-\,s}},
 \label{eq:jet-ht4} 
\end{eqnarray}
where $\hat s=x^+x^-s$ and, in view of the numerous brute force assumptions made,
we introduced an adjustable parameter $K_{\rm HT}$ which controls the 
magnitude of the HT corrections in our approach.

Including also higher power corrections, accounting for the GW decomposition
of hadron wave functions, and expressing multiparton GPDs via single parton 
ones (thereby neglecting parton-parton correlations at the $Q_0^2$ scale),
\begin{eqnarray}
F^{(n)}_{I\underset{n}{\underbrace{gg...}}/h(i)}(x,x_{g_1},...,
Q^2,Q^2_{g_1},...,\vec \Delta_{g_1},...) \nonumber \\
\simeq\int \!d^2 b_I\; G^{\rm scr}_{I/h(i)}(x,b_I,Q^2) \nonumber \\
\times \prod _{i=1}^n
G^{\rm scr}_{g/h(i)}(x_{g_i},|\vec b_I+\vec \Delta_{g_i}|,Q^2_{g_i})\,,
 \label{eq:ngpd-fact} 
\end{eqnarray}
considering  rescatterings on  soft gluon pairs both from the projectile and
from the target (taking into account that those become significant in 
different parts of the kinematic space), we finally get
\begin{eqnarray}
\frac{d\sigma_{hp}^{\rm jet}(s,p_{{\rm t}})}{dp_{{\rm t}}^{2}}
\simeq\int \!d^2b \left\{K \! \int \!\!d^2b' \sum_{i,j}C^{(i)}_{h}C^{(j)}_{p}\right. \nonumber \\
\times \int\! dx^{+}dx^{-}\sum_{I,J}
G_{I/h(i)}^{\rm scr}(x^{+}+\tilde x^{+},b',\mu_{{\rm F}}^{2})\nonumber \\
\times \left.
G_{J/p(j)}^{\rm scr}(x^{-}+\tilde x^{-},|\vec b-\vec b'|,\mu_{{\rm F}}^{2})\,
\frac{2\,d\sigma_{IJ}^{2\rightarrow2}}
{dp_{{\rm t}}^{2}} \right\},
 \label{eq:jet-ht-all} 
\end{eqnarray}
where 
\begin{eqnarray}
\tilde x^{+}=x^{+}\,
\frac{K_{\rm HT}\,C_I\,\pi^2\,\alpha_{\rm s}(\mu_{\rm R}^2)}{|\hat t|}\nonumber \\
\times \left.
x_g^+G_{g/h(i)}^{\rm scr}(x_g^+,b',Q_0^{2})\right|_{x_g^+=\frac{Q_0^2}{x^-\,s}} 
 \label{eq:xptild} \\
\tilde x^{-}=x^{-}\,
\frac{K_{\rm HT}\,C_J\,\pi^2\,\alpha_{\rm s}(\mu_{\rm R}^2)}{|\hat t|}\nonumber \\
\times \left.
x_g^-G_{g/p(j)}^{\rm scr}(x_g^-,|\vec b-\vec b'|,Q_0^{2})
\right|_{x_g^-=\frac{Q_0^2}{x^+\,s}}.
 \label{eq:xmtild} 
\end{eqnarray}

Like in the original approach of Refs.\  \cite{qiu04,qiu04a,qiu06}, the effect
of the considered HT corrections amounts to a shift of the LC$^{\pm}$ momentum
fractions of the active partons $I$ and $J$, participating in the hard 
scattering.
 However, in our case,
due to the assumed model for multiparton correlators, the magnitude of these
shifts $\tilde x^{\pm}$, Eqs.\ (\ref{eq:xptild}-\ref{eq:xmtild}),
depends on the collision kinematics. Consequently, the overall strength of the
HT  corrections varies with the energy and impact parameter of the collision,
 increasing in the    ``dense'' (high $s$, small $b$) limit.

Modifications similar to Eq.\ (\ref{eq:jet-ht-all}) apply to all the eikonals
describing ``semihard Pomeron'' exchanges.
For example, $\chi_{hp(ij)}^{\mathbb{P}_{{\rm sh}}}$, Eq.\ (\ref{chi-sh-gw}),
now takes the form
\begin{eqnarray}
\chi_{hp(ij)}^{\mathbb{P}_{{\rm sh}}}(s,b)=\frac{K}{2}\sum_{I,J}
\int\! d^{2}b'\int\! dx^{+}dx^{-} &&\nonumber \\
\times \int \!dp_t^2 \;
G_{I/h(i)}(x^{+}+\tilde x^{+},b',\mu_{{\rm F}}^{2})\nonumber \\
\times \;
G_{J/p(j)}(x^{-}+\tilde x^{-},|\vec b-\vec b'|,\mu_{{\rm F}}^{2}) \nonumber \\
\times \;\frac{d\sigma_{IJ}^{2\rightarrow2}(\hat s,p_{t}^{2},\mu_{\rm R})}
{dp_{{\rm t}}^{2}}\,\Theta(\mu_{F}-Q_{0})\,,
\label{chi-sh-gwn}&&
\end{eqnarray}
with  $\tilde x^{\pm}$ as in Eqs.\ (\ref{eq:xptild}-\ref{eq:xmtild}), and with $G_{I/h(i)}$ being defined neglecting
absorptive corrections [c.f.\ Eq.\ (\ref{eq:gpd-gw})]:
\begin{eqnarray}
G_{I/h(i)}(x,b,\mu_{{\rm F}}^{2})=  \sum _{I'}\int_x^1 \! dz
\nonumber \\
\times \,E_{I'I}^{{\rm QCD}}(z,Q_{0}^{2},\mu_{{\rm F}}^{2})\,
 \chi_{Ih(i)}^{\mathbb{P}_{{\rm soft}}}(s_{0}z/x,b)/x\,.
\label{eq:pdfs-noscr} 
\end{eqnarray}
Neglecting the HT corrections, i.e., setting $K_{\rm HT}=0$,
 Eqs.\ (\ref{eq:sig-2jet}) and (\ref{chi-sh-gw}) are recovered from Eqs.\ 
 (\ref{eq:jet-ht-all}) and (\ref{chi-sh-gwn}), respectively.
  
 In turn, for the  proton  structure function $F^{(p)}_{2}$ we obtain, using the same assumptions:
\begin{eqnarray}
F^{(p)}_{2}(x,Q^{2})\simeq x \sum _q e_q^2 \int\! d^{2}b \,\sum_i C^{(i)}_{p}
\nonumber \\
\times \; G^{\rm scr}_{q/p(i)}(x+\tilde x,b,Q^2)\,,
\label{eq:f2p} 
\end{eqnarray}
with
\begin{eqnarray}
\tilde x=
\frac{4K_{\rm HT}\,\pi^2\,\alpha_{\rm s}(Q^2)\,x}{3Q^2}
\nonumber \\ \times  \left.
x_gG_{g/p(i)}^{\rm scr}(x_g,b,Q_0^{2})\right|_{x_g=\frac{x\,Q_0^2}{Q^2}} .
 \label{eq:xtild-f2}
\end{eqnarray}

\begin{figure}[htb]
\centering
\includegraphics[height=2.5cm,width=0.45\textwidth]{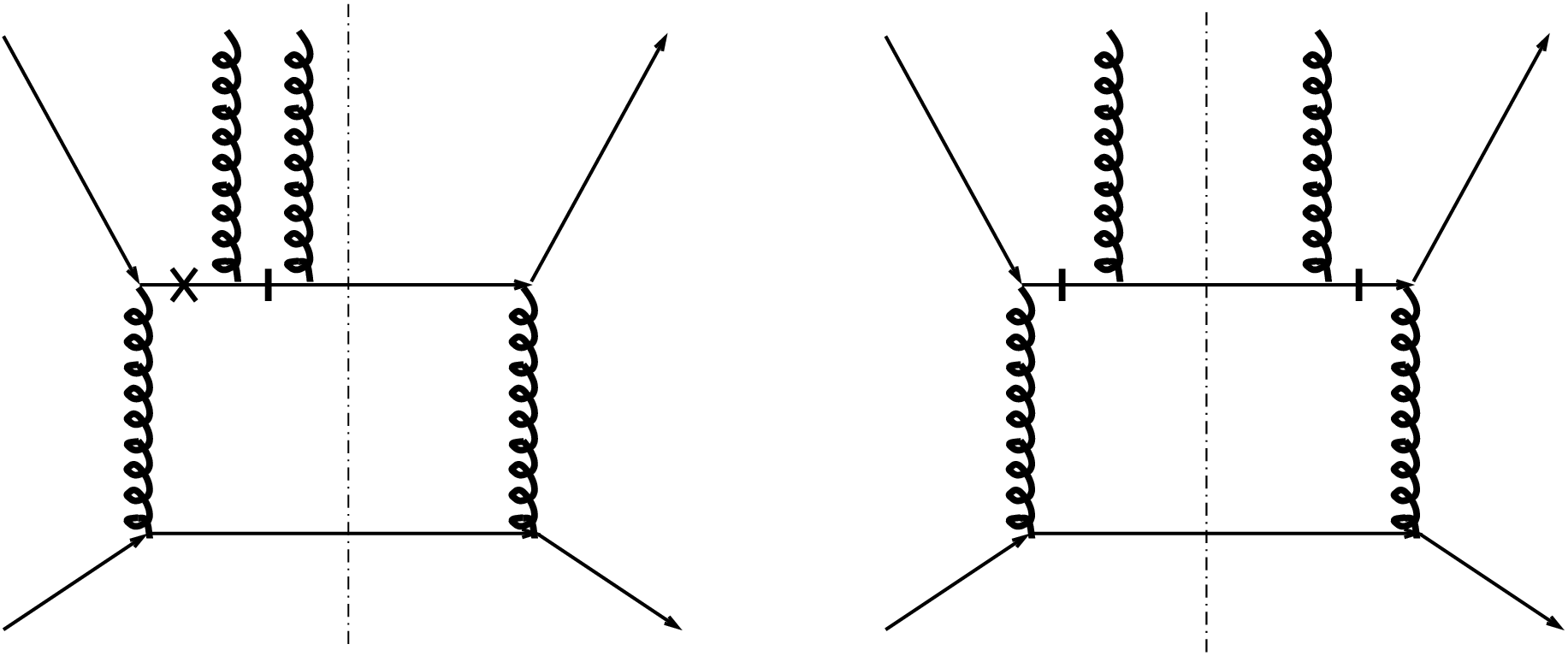}
\caption{Left: the structure of the hard ``blob'' in  Fig.\ \ref{fig:A-ht} (right),
 for leading power corrections discussed in the text,
for the case of   hard scattering of quarks of different flavors.
Right: an alternative leading power correction to the $qq'$ hard scattering,
which provides a subleading contribution
in the high energy limit. 
}
\label{fig:qq-hard}       
\end{figure}

At this point, we have to provide some arguments to support our extrapolation
of the  treatment of    \cite{qiu06} to the case of hadron-proton
 scattering. In particular,  here we can not use the $A$-enhancement argument  
 to neglect other potential HT contributions characterized by a different
 structure of the hard scattering part, compared to the one in 
      Fig.\ \ref{fig:qq-hard} (left), considered so far.
 To this end, let us remind ourselves
that the Lorentz-contraction acts differently on partons of different momenta,
in a fast-moving parton cloud of a hadron.
 While fast (large $x$) partons are confined
to a narrow ``pancake'' in the longitudinal direction, the abundant small
$x$ gluons are spread over longitudinal distances $\propto 1/(x p^+)$.
For the diagram in  Fig.\  \ref{fig:qq-hard} (left), 
corresponding to the approach of  \cite{qiu06}, the struck low-$x$ quark propagates   
over large distances $\propto 1/(x^+p^+)$ comparable to the longitudinal
size of the gluon cloud and, thus, may scatter coherently on many correlated 
  soft gluon pairs. In contrast, considering, for example, an alternative
  configuration depicted   in  Fig.\  \ref{fig:qq-hard} (right), 
  the first gluon is separated
  from the quark by the contact propagator, which implies there is 
  a very small distance between the quark and the gluon in the LC$^-$
  direction. Hence,   only a very  small portion of the gluon content 
  of the proton can be involved in that type of interaction, with the
  corresponding contribution being a subdominant one.

\section{Nuclear collisions and Monte Carlo implementation\label{MC.sec}}
The generalization of the model for the treatment of hadron-nucleus and 
nucleus-nucleus collisions is performed without introducing any additional
adjustable parameters, similarly to the case of the QGSJET-II model \cite{ost11}.
The only additional input are nucleon densities for nuclear ground states,
chosen according to the corresponding experimental 
measurements\footnote{For light nuclei, with mass number $A\leq 10$,  Gaussian
distributions are used, while densities of heavier nuclei are described by 
three parameter Wood-Saxon  distributions.} \cite{vri87}.

As discussed in \cite{ost11}, in case of $hA$ and $AA$ collisions, enhanced Pomeron
diagrams account for rescattering of intermediate partons off different
nucleons from the projectile or/and  from the target, i.e., multi-Pomeron vertices
generally couple together sequences of Pomerons and Pomeron loops connected
to different nucleons. This gives rise to a dynamical treatment of the corresponding absorptive corrections: the strength of such nonlinear effects increases with the collision energy, with the size of the colliding nuclei,
and with the ``centrality'' of the collision.

The very same tendencies hold also for the HT effects discussed in Section 
\ref{ht.sec}, for  $hA$ and $AA$ collisions. Indeed, the multiparton correlators,
Eqs.\ (\ref{eq:T-ht-q}) and (\ref{eq:T-ht-g}), involve in those cases soft gluons emitted by different nucleons. 
Describing such correlators by multiparton GPDs,
 the HT corrections  rise in the low $x$ and small $b$ limits, increasing 
 also with the size of the nuclei.

The MC procedure for generating individual inelastic scattering events is almost
identical to the one of the QGSJET-II model \cite{ost11}: starting from sampling
the impact parameter for a collision, according to the respective interaction profile, and proceeding to specifying  a ``macro-configuration'' of the event, i.e.,  defining the structure of the corresponding cut Pomeron ``net'', 
based on 
partial  cross sections for such macro-configurations. This is followed
by sharing the energy-momentum between all ``elementary'' parton cascades
(cut Pomerons) and choosing, with the appropriate weight, whether a particular
cut Pomeron involves a purely ``soft'' ($|q^2|<Q_0^2$) parton evolution or
corresponds to a semihard parton cascade. In the latter case, one generates
explicitly the initial and final state parton emission, treating the 
respective $t$- and $s$-channel parton cascades, using the DGLAP formalism.

More specifically, the initial ($t$-channel) parton emission is modeled
using a forward evolution algorithm, based on an integral representation
of the DGLAP   equations. For any parton (sub)ladder of mass squared
$\hat s$, with the ladder ``leg'' partons $I$ and $J$, characterized by virtualities $q_1^2$ and  $q_2^2$, respectively ($q_1^2=q_2^2=Q_0^2$ for first
partons in a particular perturbative cascade), one considers a successive
emission of a parton from any of the ladder ``ends'', according to the 
probability (see, e.g.\ \cite{dre01} for more details):
\begin{eqnarray}
f_{I'}(z,q^2) \propto \frac{\alpha_s(q^2)}{2\pi\,q^2}\,P_{I'I}(z)\,\Delta^{\rm S}_{I}(q_1^2,q^2)
&&\nonumber \\
\times \;\sigma_{I'J}^{{\rm QCD}}(z\hat{s},q^{2},q_2^{2})\,\Theta (1-\varepsilon-z)\,
\Theta(q^2-q_1^2) &&\nonumber \\
\times \;
\Theta(z\hat{s}/16-\max [q^{2},q_2^{2}]),\label{eq:split}
\end{eqnarray}
with $I'$,  $z$, and $q^2$   being, respectively, the type, LC (plus or
minus) momentum fraction, and virtuality of the new $t$-channel parton. 
$P_{I'I}$ are the usual (unregularized) Altarelli-Parisi splitting kernels,
$\sigma_{I'J}^{{\rm QCD}}$ is the contribution of the remaining subladder,
defined by Eq.\ (\ref{sigma-hard}), and $\Delta^{\rm S}_{I}(q_1^2,q^2)$ is the
so-called Sudakov form factor defining the probability for no parton emission
in the virtuality range $[q_1^2,q^2]$:
\begin{eqnarray}
\Delta^{\rm S}_{q}(q_1^2,q^2) = \exp \!\left[-\int _{q_1^2}^{q^2} \frac{d\tilde q^2}{\tilde q^2}\, 
\int _0^{1-\varepsilon} \!dz  \right. &&\nonumber \\
\times \left. \frac{\alpha_s(\tilde q^2)}{2\pi}\, P_{qq}(z) \right] \\
\Delta^{\rm S}_{g}(q_1^2,q^2) = \exp \!\left[-\int _{q_1^2}^{q^2} \frac{d\tilde q^2}{\tilde q^2}\, 
\int _{\varepsilon}^{1-\varepsilon} \!dz  \right. &&\nonumber \\
\times \left. \frac{\alpha_s(\tilde q^2)}{2\pi}\, \left[\frac{1}{2}P_{gg}(z) +N_f\,P_{qg}(z)\right]\right].
\label{eq:sudak-q}
\end{eqnarray}
Here $\varepsilon$ is a small enough technical ``resolution'' parameter
(we use $\varepsilon =10^{-2}$), $N_f=3$ is the number of active quark flavors, and the last $\Theta$-function in
 Eq.\ (\ref{eq:split}) is to assure that the remaining ladder is massive
 enough to allow for a parton-parton scattering with $p_t^2/4>\max [q^{2},q_2^{2}]$
 (for our choice of the factorization scale,  $\mu_{F}=p_{t}/2$).

Here comes an important difference, compared to the corresponding treatment
of \cite{ost11}. The generation of macro-configurations of collisions and the 
energy-momentum sharing procedure are quite similar to the ones of \cite{ost11},
with the only difference that the respective  ``general Pomeron'' eikonals
now contain HT corrections to hard scattering processes, as discussed in 
 Section  \ref{ht.sec}. On the other hand, when treating the $t$--channel parton
 evolution, a two step procedure is adopted. First, the corresponding parton
 emission pattern is generated based on the standard DGLAP evolution, 
 as described above, the HT
 corrections being neglected. Next, the obtained parton configuration is
 accepted with the probability defined by the ratio of the corresponding
 partial cross sections taking the HT corrections into account (i.e.,
  accounting for the kinematics-dependent shift of LC$^{\pm}$ momentum fractions
  of active partons) or neglecting them.
Otherwise, the  configuration is rejected and the procedure is repeated. 
In other words, the HT corrections to the hard scattering pattern are
 accounted for  via a rejection procedure. For brevity, we omit here the
 corresponding technical details.

In turn, the modeling of the final  ($s$-channel) parton emission follows
closely the procedure described in \cite{mar84}, imposing the final
transverse momentum
cutoff $p^2_{{t,\rm cut (f)}}=0.15$ GeV$^2$. In particular, one assures angular
 ordering of sequential $s$-channel subcascades, resulting from color coherence
effects \cite{dok80,bas83}.

As the final step, one considers  a formation of strings of color
field, stretched between constituent partons of the interacting hadrons (nuclei) or/and between the final $s$-channel partons resulting from the above-discussed
treatment of perturbative parton cascades, following the directions of the
corresponding color flows. The breakup and hadronization of such strings 
is modeled by means of a string fragmentation procedure, to be discussed elsewhere \cite{ost23}.
 Here two important comments are in order. First, the color connections
 between final partons are defined based on the $1/N_c$ approximation ($N_c$
 being the number of colors):  following the directions of the
 color and ``anticolor'' flows \cite{and83}. For each $s$-channel gluon emission, there are two alternative
 ways (taken with equal probabilities) to continue such flows, such that a
 diagram with $n$  $s$-channel gluons can produce up to $2^n$ possible patterns for
 the string configuration (see  \cite{wer23} for a recent detailed discussion). Secondly, in the hadronization procedure of the
 QGSJET-III model, inherited from the original QGSJET \cite{kal97}, one considers
 a nonperturbative splitting of final gluons into quark-antiquark pairs,
 with the strings having such (anti)quarks at the ends, without gluon ``kink''
 perturbations. Such an approach is generally inferior in quality, compared  to more advanced
 ``kinky string'' hadronization procedures  \cite{and83,dre02} 
 employed in the  PYTHIA \cite{sjo06}  and EPOS  \cite{wer06} MC generators, notably,
 regarding high $p_t$ production of relatively heavy hadrons.

\section{Selected results and discussion\label{results.sec}}
The parameters of the model have been fixed using experimental data on
total, elastic, and diffractive hadron-proton cross sections, on the proton
SF $F_2$, and on secondary hadron production   in $hp$ and $hA$ 
collisions\footnote{A comparison with experimental data on secondary hadron production and a discussion of  model parameters related to the hadronization procedure will be presented elsewhere \cite{ost23}.}, the corresponding values
being compiled in Table 1. As one can see in the Table, replacing the
 projectile proton by pion or kaon, subject to a change are only the parameters
 describing the transverse sizes of GW Fock states of
 the hadrons. 
 \begin{table*}[t]
\begin{centering}
\begin{tabular}{|p{5mm}lp{5mm}lp{5mm}llp{4mm}lp{4mm}lp{4mm}lp{4mm}ll
llllllll|}
\hline 
 {\scriptsize $\alpha_{\mathbb{P}}(0)$} & {\scriptsize $\alpha'_{\mathbb{P}}(0)$} & {\scriptsize $Q_0^2$} & {\scriptsize $r_{3\mathbb{P}}$}
  & {\scriptsize $K_{\rm HT}$} & {\scriptsize $r_{g/\mathbb{P}}$} & {\scriptsize $w_{qg}$} & {\scriptsize $\xi$}  & {\scriptsize $\beta_{p}$}   & {\scriptsize $\beta_{\pi}$}   & {\scriptsize $g_0$}
  & {\scriptsize $d_p$} & {\scriptsize $d_{\pi}$} & {\scriptsize  $d_{K}$}
    & {\scriptsize $R^2_{p(1)}$}  & {\scriptsize $R^2_{\pi (1)}$} & {\scriptsize  $R^2_{K (1)}$} 
  \tabularnewline
&  {\scriptsize GeV$^{-2}$} &  {\scriptsize GeV$^2$} &  {\scriptsize GeV$^{-1}$} &&&&& & & &
 & & & {\scriptsize GeV$^{-2}$} &  {\scriptsize GeV$^{-2}$} &  {\scriptsize GeV$^{-2}$}
  \tabularnewline
\hline 
 {\scriptsize 1.21} & {\scriptsize 0.21} & {\scriptsize 2} & {\scriptsize 0.3} &   {\scriptsize 2.5} & {\scriptsize 0.11} & {\scriptsize 0.28}  & {\scriptsize 1.5}  &
{\scriptsize 4} & {\scriptsize 2} &{\scriptsize 2} & 
{\scriptsize 0.11} & {\scriptsize 0.2} & {\scriptsize 0.28} & {\scriptsize 5.58} & {\scriptsize 1.9} & {\scriptsize 1.12}
\tabularnewline
\hline
\end{tabular}\caption{Model parameters.}
\label{Flo:param}
\par\end{centering}
 \end{table*}

\begin{figure}[htb]
\centering
\includegraphics[height=6cm,width=0.49\textwidth]{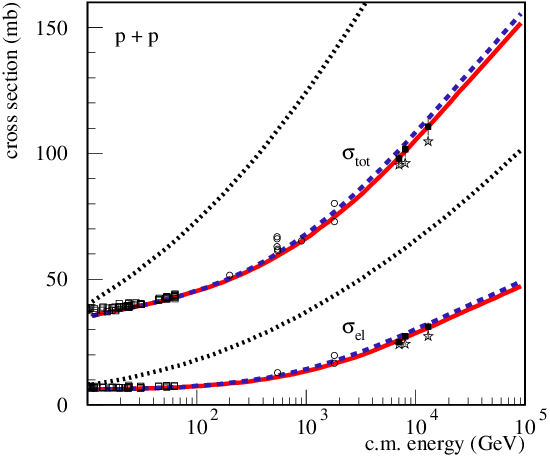}
\caption{Calculated energy-dependence of the total and elastic $pp$ cross
sections -- solid lines. The results obtained 
omitting the HT corrections  or neglecting 
all nonlinear effects are shown by dashed and dotted lines, respectively.
 The experimental data (points) are 
from Refs.\ \cite{pdg,aad23}.}
\label{fig:sig-ht}       
\end{figure}

 In Fig.\ \ref{fig:sig-ht},
 we compare the calculated energy-dependence of the total $\sigma_{pp}^{\rm tot}$
 and elastic $\sigma_{pp}^{\rm el}$ proton-proton cross sections
  to accelerator data. Additionally,
 we show by dashed lines the corresponding results obtained neglecting the
 HT corrections, i.e., setting $K_{\rm HT}=0$, while keeping all the other
  parameters unchanged. In turn, dotted lines correspond to the case of all
  nonlinear effects being neglected, i.e., setting also the triple-Pomeron
  coupling $r_{3\mathbb{P}}=0$. It is easy to see that the highest impact on the
  $\sqrt{s}$-dependence of $\sigma_{pp}^{\rm tot/el}$ is due to absorptive
  corrections generated by enhanced Pomeron diagrams. On the other hand, the
  considered HT corrections also affect somewhat the calculated 
$\sigma_{pp}^{\rm tot/el}$ at sufficiently high energies.

\begin{figure*}[t]
\centering
\includegraphics[height=5.cm,width=0.9\textwidth]{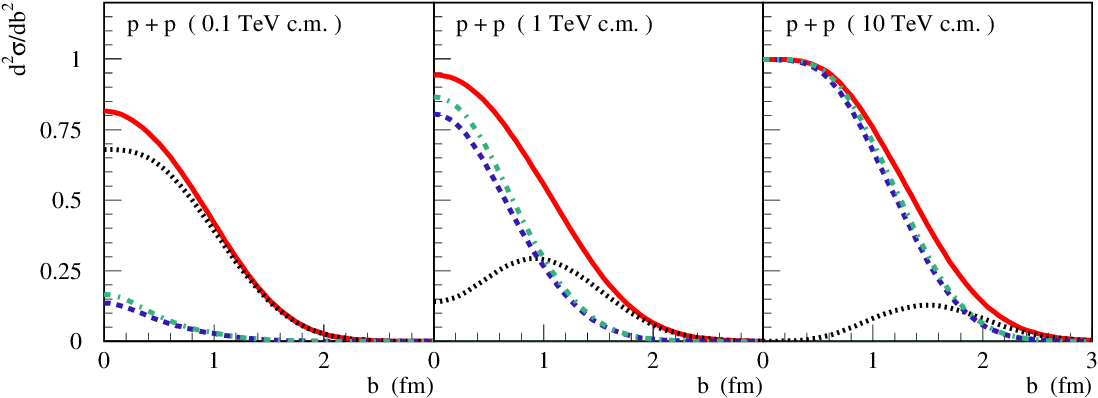}
\caption{Calculated transverse profile for the absorptive part 
of $\sigma^{\rm inel}_{pp}$
for $\sqrt{s}=10^2$ (left), $10^3$ (middle), and  $10^4$ GeV (right) 
-- solid lines.
Partial contributions of soft and semihard production processes are shown by
dotted and dashed lines, respectively. Dashed-dotted lines correspond to the latter contribution 
 calculated neglecting the HT corrections.}
\label{fig:sig-prof}       
\end{figure*}

Even such a moderate effect on  $\sigma_{pp}^{\rm tot/el}$ 
may seem somewhat surprising since the HT corrections
apply to hard scattering processes only. Such processes can be expected
to dominate relatively central (small $b$) collisions of hadrons, while having
a weak impact on the large $b$ behavior of the scattering amplitude. In that
regard, it is useful to remind oneself that in the ``semihard Pomeron'' scheme
employed here, hard parton scattering is preceded by a long enough
``soft preevolution'' \cite{dre01,ost02}, as discussed in Section \ref{hard-pom.sec}.
With increasing energy, such soft parton evolution covers a larger rapidity
interval and the corresponding transverse diffusion gives rise to a substantial
widening of the transverse profile for semihard scattering processes.
This is illustrated in Fig.\  \ref{fig:sig-prof},
 where the transverse profile for the absorptive part of the inelastic cross section,
 $d^2 \sigma^{\rm abs}_{pp}/d^2b$, for $pp$ collisions at $\sqrt{s}=10^2$, $10^3$,
and  $10^4$ GeV is plotted, together with partial contributions of purely
soft (i.e., with only soft Pomerons being cut) or semihard (with some ``real''
parton cascades entering the perturbative, $|q^2|>Q_0^2$, domain) particle production. Additionally shown by dashed-dotted lines is the semihard
contribution calculated without the HT corrections, i.e., setting $K_{\rm HT}=0$.
As is easy to see in the Figure, not only the normalization of the profile for semihard interactions rises with energy but also its width increases, corresponding to such interactions happening at larger and larger impact parameters.
Further noteworthy is the strong damping of  purely soft production processes
at small $b$ and large  $\sqrt{s}$, caused by absorptive corrections due to
virtual semihard scattering processes\footnote{Roughly speaking, the corresponding
damping factor can be interpreted as the probability to have no semihard
production at the respective $b$ values.} -- see the dotted lines in
 Fig.\  \ref{fig:sig-prof}.
 
  What may also seem surprising are the relatively similar shapes of the 
 semihard scattering profiles calculated with and without the HT corrections: dashed and dashed-dotted lines in  Fig.\   \ref{fig:sig-prof}. 
 Here we remind ourselves that those
 corrections involve soft gluon GPDs $G_{g/p}(x,b,Q^2_0)$ at small $x$. Since
 those GPDs are described by soft Pomeron asymptotics (plus the relevant
 absorptive corrections), c.f.\ Eqs.\ (\ref{eq:gpd-h}-\ref{eq:GPD-Q0}),
  the corresponding significant transverse diffusion
 gives rise to a rather large slope for these GPDs, in the low $x$ limit, which
 exceeds the one for the semihard scattering itself. As a 
 consequence, these HT corrections mainly reduce the  normalization of the   semihard interaction profile, without modifying significantly its shape.

\begin{figure*}[t]
\centering
\includegraphics[height=14.cm,width=0.8\textwidth]{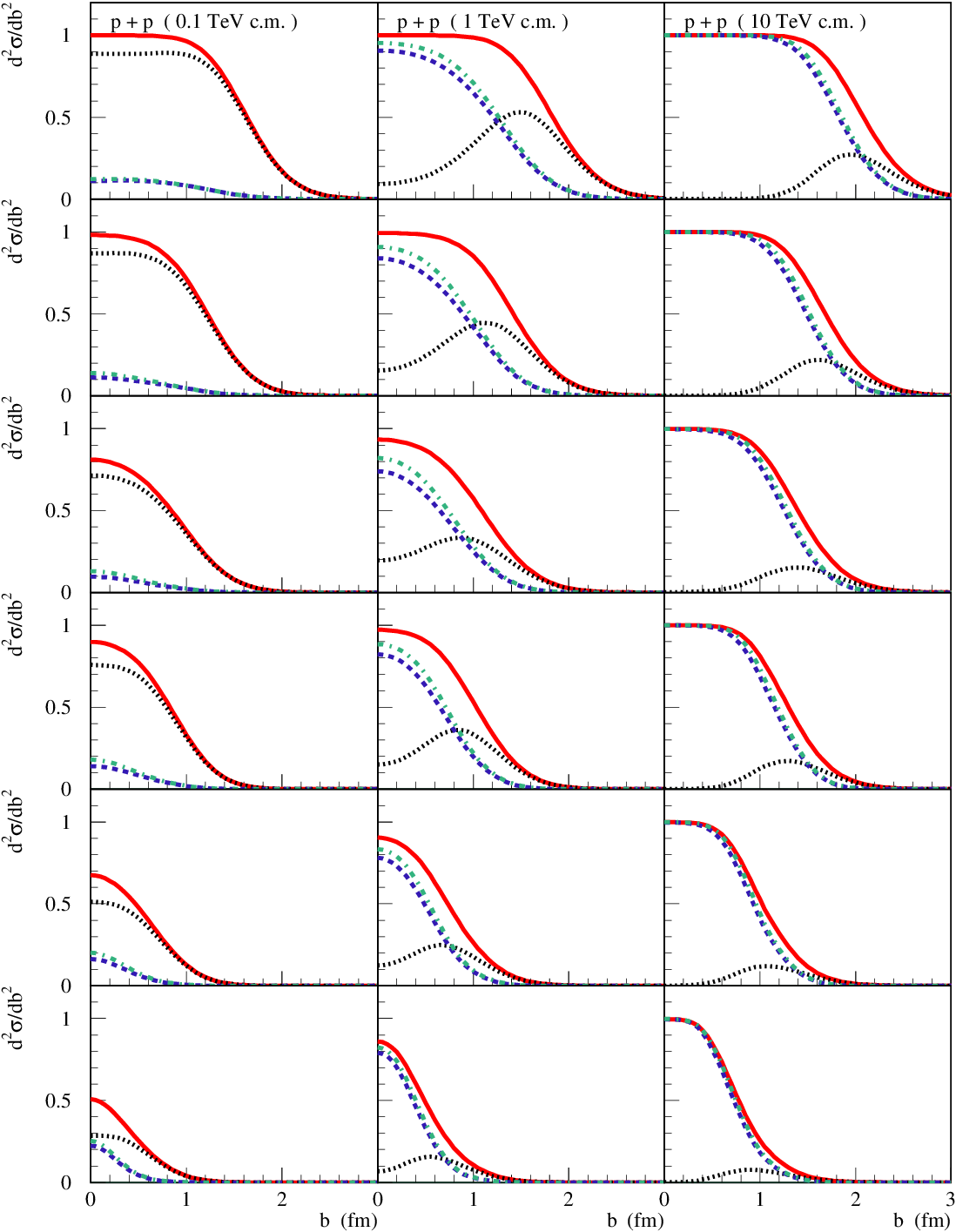}
\caption{Same as in  Fig.\  \ref{fig:sig-prof}, for particular combinations of GW 
 states  $|i\rangle$ and  $|j\rangle$
  of the projectile and target protons,
 from top to bottom: $(i,j)=(1,1)$, $(1,2)$, $(1,3)$, $(2,2)$, $(2,3)$, $(3,3)$.}
\label{fig:sig-prof-gw}       
\end{figure*}

 In relation to the model treatment of color fluctuations, it is worth
 comparing the same profiles as in Fig.\  \ref{fig:sig-prof}, for partial 
 contributions of different combinations of  GW Fock states $|i\rangle$ 
 and $|j\rangle$ of the projectile and target protons, respectively,
 as plotted in  Fig.\  \ref{fig:sig-prof-gw}. 
 The first thing to notice is the quick energy-rise
 of the relative importance of smaller size GW states: both due to increasing
 opaqueness of the corresponding profiles at small $b$ and due to a fast  transverse broadening of these profiles. The two effects are due to, respectively, the low $x$ rise of the 
 (initially small at large $x$) parton densities
 and due to the quick transverse expansion of the (initially
 compact) parton ``clouds'' of small size GW states.
  Secondly, the damping of purely
 soft hadron production is substantially stronger for larger size GW states,
 owing to their larger (integrated) parton densities. Likewise, the HT
 corrections to the calculated transverse profiles are  more significant
 for larger GW Fock states: because of   larger soft gluon densities of
 those states. 
 
\begin{figure*}[t]
\begin{centering}
\includegraphics[width=0.9\textwidth,height=6.5cm]{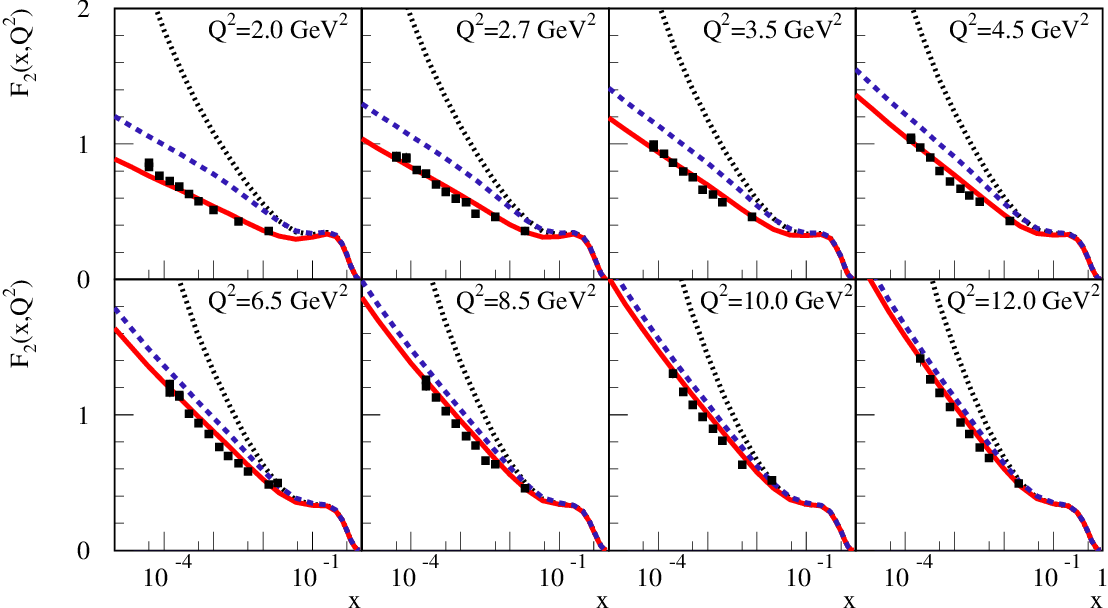}\caption{Calculated $x$-dependence 
of the  proton  structure function $F_{2}^{(p)}(x,Q^{2})$, for different $Q^{2}$,
 as indicated in the plots, compared to HERA data \cite{aar10}.\label{fig:f2}}
\par\end{centering}
\end{figure*}

\begin{figure}[htb]
\centering
\includegraphics[height=3.2cm,width=0.49\textwidth]{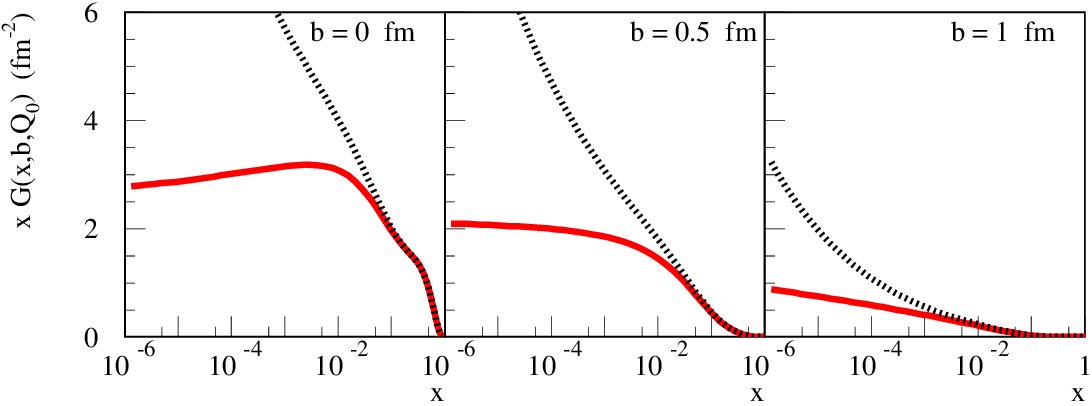}
\caption{Gluon GPD $x\,G_{g/p}(x,b,Q_0^2)$ 
for different values of $b$ (as indicated in the plots), calculated with
(solid lines) and without (dotted lines) absorptive corrections due to Pomeron-Pomeron interactions.}
\label{fig:pdfb}       
\end{figure}

 In Fig.\ \ref{fig:f2}, we compare the calculated proton SF $F_2^{(p)}$, 
 Eq.\ (\ref{eq:f2p}), to the corresponding
 HERA data, plotting also the results obtained either by suppressing the
 corresponding HT corrections (setting $K_{\rm HT}=0$) or by neglecting all
 nonlinear effects (setting also   $r_{3\mathbb{P}}=0$). As one can see
 in the Figure, the considered power corrections have a significant impact on the
 calculated proton SF $F_2$ at relatively small values of $Q^2$, with the effect
 vanishing for sufficiently high  $Q^2$. Clearly, these data constrain 
 considerably the magnitude of the HT corrections in the
 model, i.e., the value of the parameter $K_{\rm HT}$.
    On the other hand, much more serious effect is due to the taming 
 of the  low $x$ rise of proton PDFs, caused by absorptive
  corrections due to enhanced Pomeron graphs.
 To discuss the latter in more detail, it is worth considering the 
 $x$-dependence of the gluon GPD  $G_{g/p}(x,b,Q^2_0)$, calculated with and
 without such corrections, for different $b$ values \cite{ost05,ost19}, as plotted
 in Fig.\ \ref{fig:pdfb}.
 Here one clearly sees the strong $b$-dependence of nonlinear
  effects due to Pomeron-Pomeron interactions:
   while being moderate at large $b$,   such corrections damp strongly 
   the low $x$ rise of the gluon GPD
 in the $b\rightarrow 0$ limit, causing a saturation of the gluon
  density.

\begin{figure}[htb]
\centering
\includegraphics[height=8cm,width=0.49\textwidth]{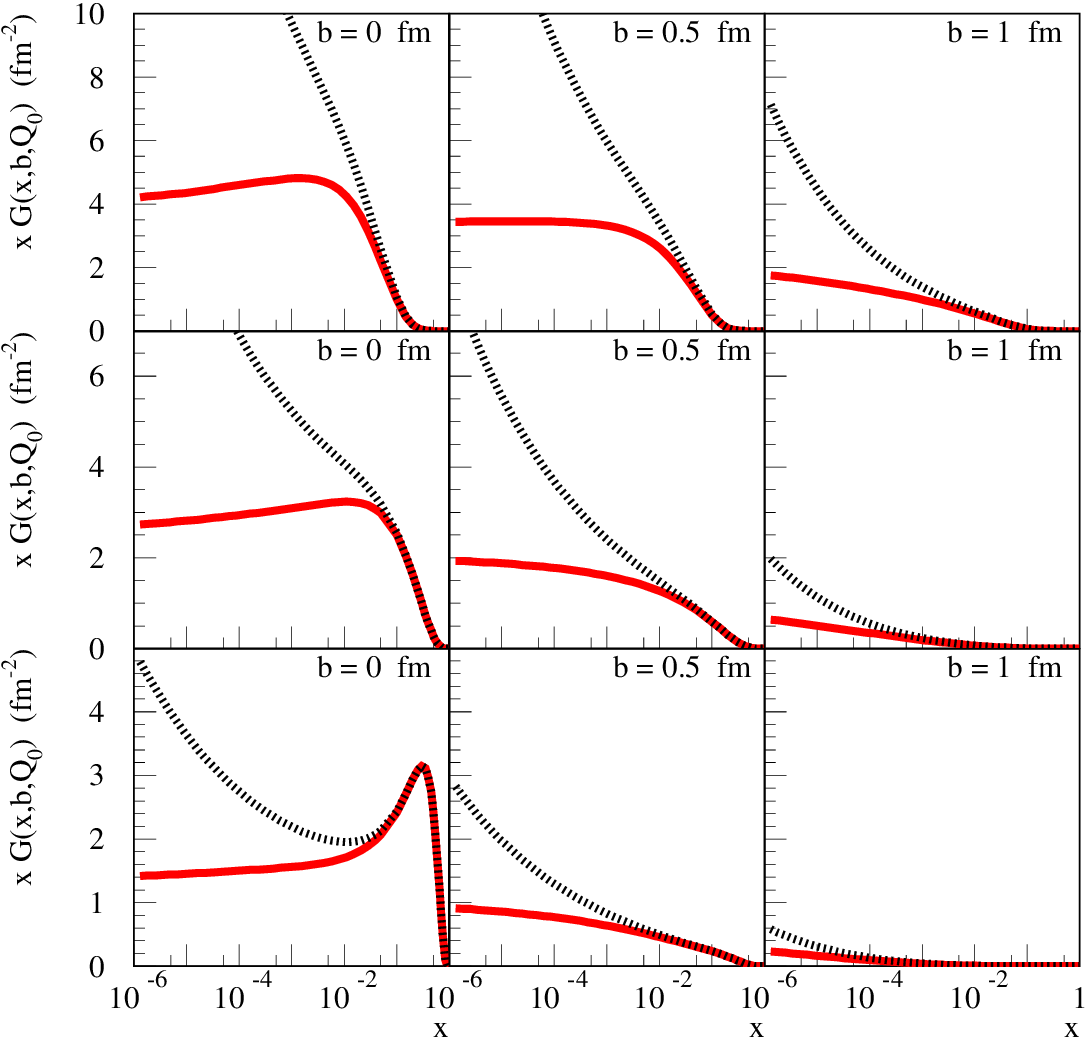}
\caption{Same as in  Fig.\  \ref{fig:pdfb}, for partial
gluon GPDs, $x\,G_{g/p(i)}(x,b,Q_0^2)$, of different
GW Fock states;
 from top to bottom: $i=1,2,3$.}
\label{fig:pdfb-gw}       
\end{figure}

   It is worth considering also the same dependence for partial gluon GPDs
   of different GW Fock states,  $G_{g/p(i)}(x,b,Q^2_0)$, as plotted in 
Fig.\  \ref{fig:pdfb-gw}. 
 Here 
  we observe the strongest impact of 
nonlinear effects on the low $x$ behavior of the partial GPDs of the largest
Fock states: as a consequence of their largest soft parton densities.

\begin{figure}[htb]
\begin{centering}
\includegraphics[width=0.49\textwidth,height=8.2cm]{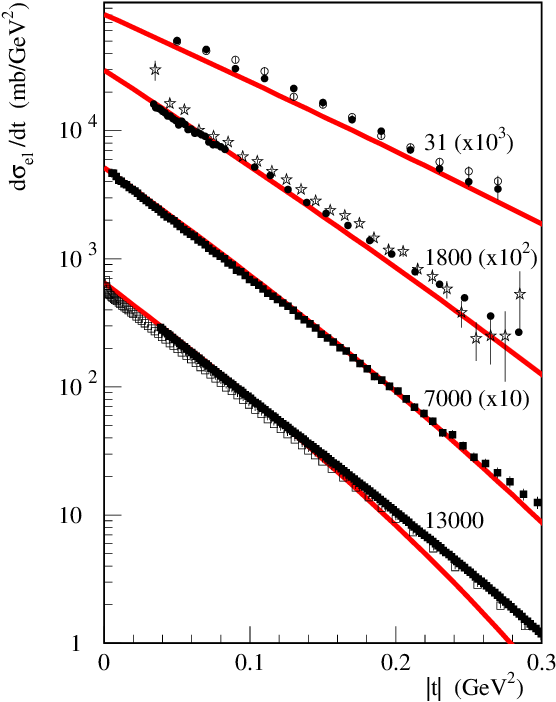}
\par\end{centering}
\caption{Calculated differential elastic proton-proton cross section
for different $\sqrt{s}$ in GeV (as indicated in the plot), compared to
experimental data (points) from Refs.~\cite{aad23,bre84,amo90,abe94b,ant13,ant19}.\label{fig:dsigel}}
\end{figure}

In  Fig.\ \ref{fig:dsigel},
 we compare the calculated differential elastic cross section
 $d\sigma^{\rm el}_{pp}/dt$ to experimental measurements. While the agreement
with the data is satisfactory for small values of $|t|$, where the bulk of the
contribution to $\sigma^{\rm el}_{pp}$ comes from, this is not the case for
larger $|t|\gtrsim 0.2$. Generally, a better description of the observed
$t$-dependence of $d\sigma^{\rm el}_{pp}/dt$ requires a more sophisticated choice
for the proton form factor (see, e.g.\ \cite{kmr18}), compared to the simple
Gaussian ansatz used in the current work. Moreover, for relatively large $|t|$,
the simple GW decomposition of the proton wave function, with $t$-independent
partial weights and profiles of GW Fock states, becomes invalid \cite{fra22}.
Since the presented model aims at describing the bulk of general hadronic
and nuclear scattering processes, such potential developments are beyond
the scope of the current study.

\begin{figure}[htb]
\centering
\includegraphics[height=5.5cm,width=0.49\textwidth]{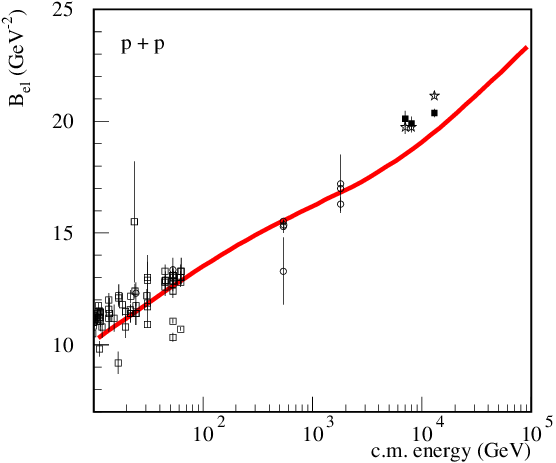}
\caption{Energy-dependence of  the calculated  forward elastic scattering slope
for $pp$ collisions, compared to experimental data (points) 
from Refs.\ \cite{pdg,aad23,ant13,ant19,ant13a,aad14,aab16}.}
\label{fig:bel}       
\end{figure}

Additionally, in  Fig.\ \ref{fig:bel},
 we compare to experimental data the calculated $\sqrt{s}$-dependence
  of the elastic scattering slope 
$B^{\rm el}_{pp}=d\ln \left.d\sigma^{\rm el}_{pp}/dt\right|_{t=0}$,
which quantifies the energy-dependence of the average squared impact parameter
$\langle b^2\rangle$ for $pp$ collisions. Clearly, the model fails to reproduce
the rather large values of $B^{\rm el}_{pp}$, reported by the TOTEM and ATLAS
collaborations at LHC energies. On the one side, this may be related to 
additional physics mechanisms missing in the model,
 like the pion loop contributions to the 
Pomeron Regge trajectory \cite{ans72} (see  \cite{kmr15} for a recent discussion).
On the other hand, it may indicate a certain deficiency of the
 treatment of color fluctuations in the model: as one can see in 
  Fig.\   \ref{fig:bel}, the increasing relative
importance of smaller size GW states slows down the energy rise
of  $B^{\rm el}_{pp}$ at $\sqrt{s}\gtrsim 1$ TeV.

\begin{figure}[htb]
\centering
\includegraphics[height=6.cm,width=0.49\textwidth]{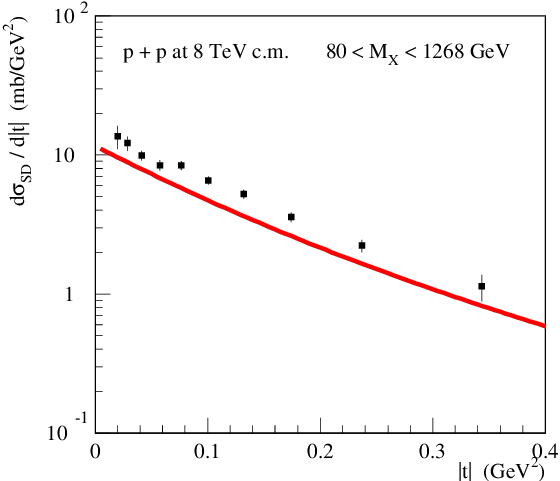}
\caption{Calculated    $d\sigma^{\rm SD}_{pp}/dt$ at $\sqrt{s}=8$ TeV,
for  $10^{-4}<\xi<10^{-1.6}$, compared to ATLAS data (points) \cite{aad22}.}
\label{fig:sig-sd}       
\end{figure}

Let us now come to the model predictions for the inelastic diffraction. 
In view of a certain tension between different LHC results on the cross section
for high mass diffraction, discussed previously in \cite{ost14,kmr14},
of significant importance is the recent measurement by the
ATLAS  experiment of the  differential 
single diffractive (SD) cross section, $d\sigma^{\rm SD}_{pp}/dt$,
at $\sqrt{s}=8$ TeV,  using the Roman Pot technique \cite{aad22}. Comparing in
Fig.\ \ref{fig:sig-sd} the calculated\footnote{The calculation involves a  Fourier transform to impact parameter space, in order to account for all the relevant
absorptive corrections (see, e.g.\ \cite{lun09}).}  $d\sigma^{\rm SD}_{pp}/dt$,
 for the experimental event selection $10^{-4}<\xi=M_X^2/s<10^{-1.6}$, $M_X$
being the  diffractive state mass, with the ATLAS data,   we observe a rather good agreement for the obtained $t$-dependence, while the magnitude of 
the calculated cross section (1.4 mb)
  is  $\sim 25$\% below the measured value, $1.88\pm 0.15$ mb.\footnote{Taking into account that an
  additional $\sim 40$\% contribution to the measured cross section, for the selected $\xi$-range, comes from  nondiffractive  production \cite{ost23} (so-called  random rapidity gaps \cite{kmr10}),
   the total  SD-like event rate predicted actually exceeds the measurement
   by $\sim 20$\%.} 
 
\begin{figure}[htb]
\centering
\includegraphics[height=6cm,width=0.49\textwidth]{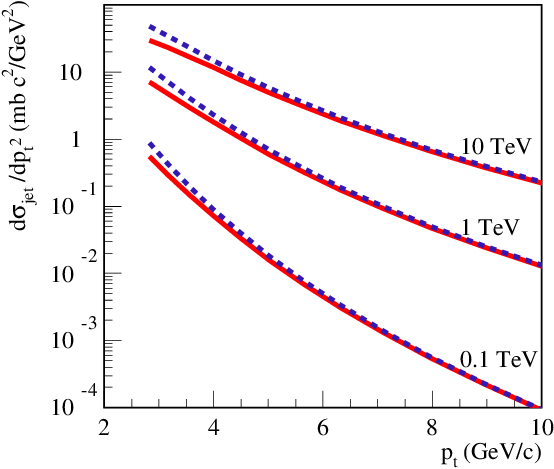}
\caption{Transverse momentum spectrum for (mini)jet production in $pp$ collisions
at   $\sqrt{s}=10^2$, $10^3$, and  $10^4$ GeV, as indicated in the plot,
calculated with and without HT corrections -- solid and dashed lines,
respectively.}
\label{fig:ptjet-ht}       
\end{figure}

\begin{figure}[htb]
\centering
\includegraphics[height=6cm,width=0.49\textwidth]{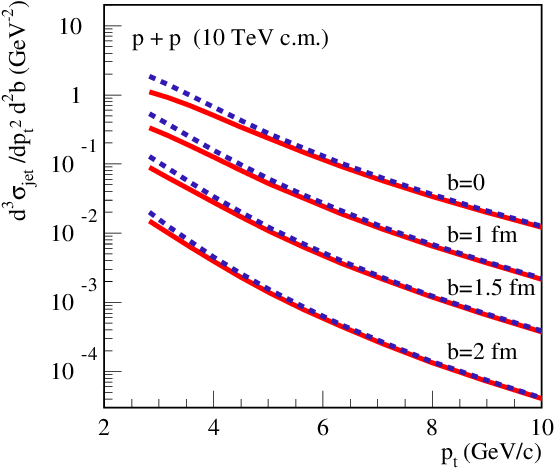}
\caption{Transverse momentum spectrum for (mini)jet production in $pp$ collisions
at  $\sqrt{s}=10^4$ GeV,
for different impact parameters, 
as indicated in the plot,
calculated with and without HT corrections -- solid and dashed lines,
respectively.}
\label{fig:ptjet-ht-b}       
\end{figure}

Finally, to illustrate the impact of the considered HT effects on secondary
hadron production, we compare in   Fig.\ \ref{fig:ptjet-ht}  
 the $p_t$-dependence of the
(mini-)jet production cross section, $d\sigma^{\rm jet}_{pp}(s,p_t)/dp_t^2$,
 at  $\sqrt{s}=10^2$, $10^3$, and  $10^4$ GeV, as calculated with and without the HT corrections, i.e., using Eqs.\ (\ref{eq:jet-ht-all}) and 
(\ref{eq:sig-2jet}), respectively. It is easy to see that such corrections reduce  
 $d\sigma^{\rm jet}_{pp}(s,p_t)/dp_t^2$ considerably in the small $p_t$ limit,
 with the suppression becoming more and more significant at higher energies.
 On the other hand, as expected, the effect vanishes for sufficiently high $p_t$.
 Additionally, in Fig.\ \ref{fig:ptjet-ht-b},
 we consider contributions of different impact parameters 
  to these jet spectra, i.e., we plot
 $d^3\sigma^{\rm jet}_{pp}(s,p_t)/dp_t^2/d^2b$ defined by the expression
 in the curly brackets in Eq.\ (\ref{eq:jet-ht-all}).
  The impact of the considered HT corrections on low $p_t$
 (mini-)jet production becomes maximal for $b\simeq 0$, while decreasing
 slowly with the increase of $b$.
 
\section{Conclusions\label{concl.sec}}
We presented here a new model for high energy hadronic scattering, QGSJET-III,
discussing in some detail its underlying theoretical mechanisms.
In particular, a considerable attention has been devoted to a phenomenological 
treatment of dynamical power corrections to hard parton scattering 
processes, based on the approach of Refs.\ \cite{qiu04,qiu06}: with the respective
contributions being  related to coherent rescattering of $s$-channel partons
on soft gluon pairs emitted by the colliding hadrons (nuclei). Modeling the
corresponding multiparton correlators as multiparton GPDs, we developed a 
dynamical scheme: with the strength of the HT effects increasing both in the  very high energy and small impact parameter limits.

Additionally, we  discussed in some detail the model implementation of color fluctuation
 effects: based on a decomposition of hadron wave functions in
a number of GW Fock states characterized by different transverse sizes and
different parton densities.

Selected model results regarding the energy dependence of the total and elastic
proton-proton cross sections, the $x$-dependence of the proton SF $F_2$, the
$x$- and $b$-dependence of the gluon GPD $G_{g/p}(x,b,Q^2_0)$, and the
 $t$-dependence of single-diffractive $pp$ cross section have been presented
 and the impact of various nonlinear corrections to the interaction  dynamics
 has been investigated. Overall, the most important feature of the model
 is the microscopic treatment of nonlinear effects due to Pomeron-Pomeron
 interactions, which is  inherited from
  the previous model version, QGSJET-II. On the other hand, the developed
  phenomenological treatment of HT effects serves the principal goal:
  taming the steep rise of the (mini-)jet production in the small $p_t$ limit,
  thereby reducing considerably  the sensitivity of the model results
   to the choice of the  ``infrared'' cutoff $Q^2_0$, as discussed previously
   in \cite{ost19}.
   
   The description of secondary hadron production and applications of the 
   model to modeling the development of CR-induced extensive air showers
   will be discussed elsewhere \cite{ost23}.

\subsection*{Acknowledgments}

This work was  supported by  Deutsche Forschungsgemeinschaft 
(project number 465275045).

\end{document}